\documentclass[journal]{IEEEtran}
\usepackage{amssymb,amsmath,graphicx,subfigure}
\usepackage{float}
\usepackage{multirow}
\usepackage{multicol}
\usepackage{hyperref}
\newtheorem{definition}{Definition}
\newtheorem{theorem}{Theorem}
\newtheorem{lemma}{Lemma}
\newtheorem{proof}{Proof}
\newtheorem{claim}{Claim}{\bfseries}{}
\usepackage{cite}
\usepackage{xcolor}

\newcommand{\etc}{{\em etc.}}
\newcommand{\etal}{{\em\,et al.\ }}
\newcommand{\eg}{e.g.\ }
\newcommand{\ie}{i.e.\ }

\ifCLASSINFOpdf
\else
\fi
\begin{document}
%
\title{Privacy-Preserving Electronic Ticket Scheme with Attribute-based 
Credentials}
%
%
%
\author{Jinguang~Han,~\IEEEmembership{Member,~IEEE,}
        Liqun~Chen,~\IEEEmembership{Member,~IEEE,}
        Steve~Schneider, Helen~Treharne and Steve~Wesemeyer
     \thanks{J. Han, L. Chen, S. Schneider, H. Treharne and S. Wesemeyer are  
     with the Surrey Centre for Cyber Security, Department of Computer Science, 
     University of Surrey, Guildford, Surrey, GU2 7XH, United Kingdom\protect\\ 
 E-mail: \{j.han, liqun.chen, s.schneider, h.treharne, s.wesemeyer\}@surrey.ac. uk}
}
%
%

\markboth{}%
{}
%



\maketitle

\begin{abstract} Electronic tickets (e-tickets) are electronic versions of paper tickets, which
enable users to access intended services and improve services' efficiency.
However, privacy may be a concern of e-ticket users. In this paper, a
privacy-preserving electronic ticket  scheme with attribute-based credentials is
proposed to protect users' privacy and facilitate ticketing based on a user's
attributes. Our proposed scheme makes the following contributions: (1) users can
buy different tickets from ticket sellers without releasing their exact
attributes; (2) two tickets of the same user cannot be linked; (3) a ticket
cannot be transferred to another user; (4) a ticket cannot be double spent; (5)
the security of the proposed scheme is formally proven and reduced to well-known
($q$-strong Diffie-Hellman) complexity assumption; (6) the scheme has been
implemented and its performance empirically evaluated. To the best of our
knowledge, our privacy-preserving attribute-based e-ticket scheme is the first
one providing these five features. Application areas of our scheme include event
or transport tickets where users must convince  ticket sellers that their
attributes (e.g. age, profession, location) satisfy the ticket price policies to
buy  discounted tickets. More generally, our scheme can be used in any system
where access to services is only dependent on a user's attributes (or
entitlements) but not their identities.
\end{abstract}

\begin{IEEEkeywords}
Anonymity, Attribute-based Credentials, Privacy-enhanced Authentication 
\end{IEEEkeywords}

%
\IEEEpeerreviewmaketitle

\section{Introduction}\label{sec:introduction}
\IEEEPARstart{ D}{ue} to their flexibility and portability, electronic ticket
(e-ticket) systems have been extensively investigated by both industry
\cite{us,uk,uk-rail} and the academic research communities
\cite{mdnd:et,mmj:et,amm:et}. E-tickets are attractive to transport operators as
well as customers because they can reduce paper costs (tickets can be stored on
a hand-held device) and improve customer experience (tickets can be purchased
and delivered any time and anywhere). However, the use of e-tickets also raises
many questions regarding the privacy of its users due to the possibility of
linking different e-ticket transactions to a particular user --in contrast to
anonymous paper-based tickets-- and thus potentially revealing private
information, \eg working patterns, likely places of work, \etc

Therefore, an important area of research is the design of e-ticket systems which
preserve customer privacy and, moreover, can be formally proven to be secure.
Anonymous authentication, which enables users to authenticate without revealing
their identities, has been used to protect a user's privacy in many
privacy-preserving e-ticket schemes
\cite{hbcd:et2006,altdbg:et,lk:et2003,mdnd:et,i:et2014,kwg:et2013}. However,
many of these schemes were not formally proven to be secure. %
Notable exceptions are those proposed by Arfaoui {\em et al.} \cite{altdbg:et}
and Rupp {\em et al.} \cite{rhbp:p4ro}. Arfaoui {\em et al.} \cite{altdbg:et}
formally defined their security models for e-ticket schemes, including
unforgeability, unlinkability and non-repudiation, but the authors only provided
a very high-level proof. Rupp {\em et al.} \cite{rhbp:p4ro} formalised their
security models of privacy-preserving pre-payments with refunds schemes
including transportation authority security and user privacy but the security
proof of their scheme was again at a high level. 
%
Another requirement of a realistic e-ticket systems is the support for different
tickets based on a user's attributes (\eg age, location, disability, profession,
\etc), \ie to offer discounts for, say, students or disabled passengers.
However, if not implemented carefully, there is a risk that such a ticket system
reveal more information about a user than necessary when purchasing or
validating tickets. For example, a student buying a discounted student ticket
may end up revealing the university at which she is enrolled and, depending on
the student card, even her birthday neither of which is relevant to obtaining
the student discount. The minimum proof required is that she can demonstrate
that she is a legitimate student. Similarly, a disable passenger might need to
reveal more details about his disability to the ticket issuer or verifier than
necessary for purchasing or verifying a ticket. Gudymenko \cite{i:et2014} and
Kerschbaum {\em et al.} \cite{kwg:et2013} addressed this issue, but their
schemes were not proven formally.

Transport operators are naturally concerned about fraudulent use of e-tickets
due to the easy with which they can be copied. Double spend  or more generally
overspend detection, \ie the process of determining whether a ticket has been
used too many times, is therefore also an important feature that an e-ticket
scheme should support.

To address the above requirements, this paper proposes a new privacy-preserving
e-ticket scheme using attribute-based credentials which supports issuing
different tickets depending on a user's attributes. Our scheme protects an
honest user's privacy while allowing for the de-anonymisation of users who try
to use their tickets more than once (double spend detection). %
It is also a general e-ticket system and 
can be used in various application scenarios including: 

\begin{itemize} %
\item mobility as a service transport tickets (\eg rail, bus, \etc) where age,
disability, profession, affiliation, \etc might determine the prices of tickets;
\item one-off token for Internet services (\eg print service, download service
for multimedia, \etc) where age, affiliation, membership might determine the
service/access level;%
\item e-Voting where age, nationality, voting district, \etc might determine 
the voting ballot that should be issued;
\item event tickets (\eg concert, tourist attractions, conferences, \etc) where
age, affiliation, disability, \etc  might determine the ticket price/access
rights.%

\end{itemize}


\subsection{Contributions}%
In this paper, we propose a new attribute-based e-ticket scheme.

The main contributions of our scheme are: (1) {\bf\em Attribute-based
Ticketing:} users can buy different tickets depending on their signed attributes
without releasing their exact details; (2) {\bf\em Unlinkability:} two tickets
of the same user cannot been linked ; (3) {\bf\em Untransferability: } a ticket
can only be used by the ticket holder and cannot be transferred to another user;
(4)  {\bf\em Double Spend Detection:} a ticket cannot be double spent and the
identities of users who try can be revealed; (5)  {\bf\em Formal Security
Proof:} the security of the proposed scheme is formally proven and reduced to
the well-know $q$-strong Diffie-Hellman complexity assumption. (6) {\bf\em
Performance Evaluation:} the performance of our scheme has been measured on both
Android and PC platforms%

The novelty of the scheme is that it combines and extends Camenisch\etal's
set membership and range proof scheme\cite{ccs:sr2008} allowing a user to prove
that her attributes are in some sets or ranges simultaneously without revealing
the exact value of the attributes. Our scheme thus offers a natural as well as
flexible way of representing user attributes, \eg to obtain an age based
discount, a user would expect to prove that her age is in a certain range, while
for a disability discount, she would want to demonstrate her impairment is
contained within the set of recognised disabilities. Furthermore, the user
attributes are additionally signed by a trusted third party thereby allowing a
user's claimed attributes to also be verified. This is different to Camenisch's
approach to attribute verification and more suited to our application domain.



\subsection{Related Work}%
Mut-Puigserver {\em et al.} \cite{mmj:et} surveyed numerous e-ticket systems and
summarised their various functional requirements (\eg expiry date, portability,
flexibility, \etc) and security requirements (\eg integrity, authentication,
fairness, non-overspending,  anonymity, transferability, unlinkability, \etc).
E-ticket schemes are classified into different types: transferable tickets
\cite{hbcd:et2006,amm:et}, untransferable tickets \cite{iata:nontransf,mdnd:et},
multi-use tickets \cite{mdnd:et,mmj:et} and single-use tickets
\cite{mdnd:et,hbcd:et2006,amm:et,pc:et1997}.  Our scheme falls into the
untransferable, single-use tickets categories while providing anonymity,
unlinkability, non-overspending and flexibility.

We now compare our scheme with a number of other schemes. In these schemes,
blind signatures \cite{c:bs}, group signatures \cite{ch:gs}, anonymous
credentials \cite{c:ac}  and pseudonyms \cite{ch:gs,lrsw:ps1999} were used to
protect user privacy.

{\em E-Ticket Schemes from Blind Signatures.} In a blind signature scheme, a
user can obtain a signature on a message without the signer knowing the content.
Based on the blind signature scheme proposed by Chaum  \cite{c:bs}, Fan and Lei
\cite{fl:et1998} proposed an e-ticket system for voting in which each voter can
vote in different elections using only one ticket. Song and Korba
\cite{lk:et2003} proposed an e-ticket scheme to protect users' privacy and
provide non-repudiation in pay-\mbox{TV} systems. Quercia and Hailes
\cite{qh:et2005} proposed an e-ticket scheme for mobile transactions using
Chaum's blind signature scheme \cite{c:bs} to generate both limited-use and
unlimited-use tickets. Rupp\etal\cite{rbhp:p4r,rhbp:p4ro} proposed
privacy-preserving pre-payments with refunds schemes derived from Chaum's scheme
\cite{b:bs} and Boneh\etal's short signature scheme \cite{bls:s}. In their
scheme, trip authorisation tokens were generated using Chaum's blind signatures,
while Boneh\etal's short signature scheme was used to implement the
privacy-preserving aggregation of refunds. Milutinovice {\em et al.}
\cite{mdnd:et} proposed an e-ticket scheme which combines the partial blind
signature scheme proposed by Abe\etal\cite{ao:pbs2000}, Pedersen's secret
sharing commitment scheme \cite{p:com1991} and Camenisch\etal's anonymous
credential scheme \cite{cl:ac2001} to protect user privacy. All these schemes
can protect user privacy and ticket unlinkability, but, unlike our scheme, they
do not support de-anonymisation after double spending nor ticket
untransferability.

{\em E-Ticket Schemes from Group Signatures.} %
A group signature enables a user to sign  a message on behalf of the group
without exposing his identity, while the group manager can release the identity
of the real signer. Nakanishi {\em et al.}  \cite{nhs:ec1999} proposed an
electronic coupon (e-coupon) scheme where the group signature scheme
\cite{cs:gs1997} was used to provide anonymity and unlinkability. Vives-Guasch
\cite{ajmm:et2010} proposed an automatic fare collection (AFC) system in which
the group signature scheme proposed by Boneh\etal\cite{bs:gs2004} was used to
provide unlinkability and revocable anonymity. These schemes can implement
anonymity, de-anonymity, ticket unlinkability and ticket untransferability, but,
unlike our scheme, they  do not support privacy-preserving attribute-based
ticketing. While Gudymenko in \cite{i:et2014} addressed user privacy as well as
differently priced tickets in his e-ticket scheme and used group signatures to
make tickets unlinkable, no formal security models and security proofs were
presented.

{\em E-Ticket Schemes from Anonymous Credentials.} %
In an anonymous credential scheme, a user can prove to a verifier that she has
obtained a credential without releasing any other information. Heydt-Benjamin
{\em et al.} \cite{hbcd:et2006} used anonymous credentials, e-cash and proxy
re-encryption schemes to enhance the security and privacy of their public
transport e-ticket systems. %
Arfaoui {\em et al.} \cite{altdbg:et}  modified the signature scheme proposed
Boneh\etal in \cite{bb:ss2004}  to eliminate expensive pairing operations in the
verification phase, and then proposed a privacy-preserving near field
communication (NFC) mobile ticket (m-ticket) system by combining their modified
signature with the anonymous credential scheme proposed by
Camenisch\etal\cite{cl:ac2004}. In their scheme, a user can anonymously use an
m-ticket at most $k$ times, otherwise  the user is revoked by the revocation
authority. These schemes can implement anonymity, ticket unlinkability as well
as ticket untransferability, but, unlike our scheme, do not support
privacy-preserving attribute-based ticketing. Additionally, the security of
these schemes was not formally proven.

{\em E-Ticket Schemes from Pseudonyms.} %
Pseudonyms allow a user to interact with multiple organisations anonymously and
potentially without linkability. %
Fujimura and Nakajima\cite{fn:et1998} proposed a general-purpose e-ticket
framework where anonymity was achieved by using pseudonym schemes
\cite{dfty:ec1997,ggmm:anoy1997}. Jorns {\em et al.} \cite{jjq:et2007} proposed
a pseudonym scheme which could be implemented on constrained devices, and then
used it  to protect users' privacy in e-ticket systems. Kuntze and Schmidt
\cite{ks:et2007} proposed a scheme to generate pseudonym tickets by using the
identities embedded in attestation identity keys (AIKs) certified by  the
privacy certificate authority (PCA). Vives-Guasch {\em et al.}
\cite{ammj:et2012} proposed a light-weight e-ticket scheme using pseudonyms
which also addressed exculpability (\ie a service provider cannot falsely accuse
a user of having overspent her ticket, and the user is able to demonstrate that
she has already validated the ticket before using it) and reusability (\ie a
ticket can be used a predefined number of times). In \cite{ammj:et2012},
pseudonyms were used to provide unlinkability of users' transactions. Kerschbaum
{\em et al.} \cite{kwg:et2013} considered the privacy-preserving billing issue
in e-ticket schemes and applied pseudonyms to provide unlinkability of user
transactions. These schemes can implement anonymity, ticket unlinkability as
well as ticket untransferability, but, unlike our scheme, they do not support
privacy-preserving attribute-based ticketing. Furthermore, the security of these
schemes was not formally proven.

{\em E-Tickets from Special Devices.} %
There are other e-ticket schemes  designed around special devices, including
personal trusted device (PTD) \cite{ccj:et2007}, trusted platform module (TPM)
\cite{ks:et2007}, mobile handsets \cite{lll:et2016}, \etc%
 Unlike our scheme, these schemes require special devices and do not enable
de-anonymisation after double spending a ticket nor do they support
privacy-preserving attribute-based ticketing.

\subsection{Organisation} The remainder of this paper is organised as follows.
In Section~\ref{prelim}, the preliminaries used throughout this paper are
described. The construction and security analysis of our scheme are
presented in Section~\ref{const}  and Section \ref{secu}, respectively. %
In Section~\ref{sec:benchmarks}, the performance of our scheme is
evaluated. Finally, Section~\ref{conc} presents our conclusions and future work.

\section{Preliminaries}\label{prelim}%
In this section, the formal concepts and notation used throughout this
paper are introduced. The most important notation is summarised in Table
\ref{tab:1}.

\begin{table}[!t]\caption{Notation}\label{tab:1}
\centering
\begin{tabular}{c|l }
\hline
          $1^{\ell}$                &                A security number\\
          $\epsilon(\ell)$                        &                   A negligible function in $\ell$\\  
          {\sf CA}                  &                  A central authority\\
          {\sf S}                     &                   A ticket seller\\
          {\sf U}                    &                    A user\\
          {\sf V}                    &                      A ticket verifier\\
          $H$                        &         A cryptographic hash function\\
          $\mathbb{P}$        &                         A universal set of 
          ticket policies\\
          $\mathbb{P}_{U}$  &         The policies satisfied by  {\sf U}\\
          $\mathbb{R}_{j}$    &   The $j$-th range policy\\
          $\mathbb{S}_{i}$     & The $i$-th set policy\\
          $I_{i_{j}}$        & The $j$-th item in $\mathbb{S}_{i}$\\
          $\sigma_{S}$        &                    A credential of {\sf S}\\
          $\sigma_{U}$       &                    A credential of {\sf U}\\
          $A_{U}$                &                       The attributes of {\sf U}\\
          $ID_{U}$               &                      The identity of {\sf U}\\
          $ID_{S}$               &                    The identity of {\sf S}\\
          \mbox{PoK}          &                       Proof of knowledge\\
          $Ps_{U}$              &                   A pseudonym of {\sf U}\\
          $Serv$                 &                    The services requested by {\sf U}\\
          $VP_X$                    &                A validity period for X\\
           $MSK$                  &              The master secret key of the system\\
          $params$             &                 The public parameters of the system\\
          $Price$                &             The price of a ticket\\
           $Ticket_{U}$         &                  A ticket of {\sf U}\\
          $Trans_{T}$        &                 A proof transcript of  the ticket $Ticket_{U}$ \\
          $\mathcal{KG}(1^{\ell})$ & A secret-public key pair generation algorithm\\
          $\mathcal{BG}(1^{\ell})$ & A bilinear group generator\\
                           $x\stackrel{R}{\leftarrow}X$      &                   $x$ is randomly selected from the set $X$\\
          $A(x)\rightarrow y$                  &                          $y$ is obtained by running the algorithm $A(\cdot)$ \\
                                                        &  with input $x$\\
          $A_{U}\models I_{i_{j}}$ &  $A_{U}$ satisfies the item $ I_{i_{j}}$\\
          $(SK_{S},PK_{S})$ &              A secret-public key pair of {\sf S}\\
          $(SK_{U},PK_{U})$ &              A secret-public key pair of {\sf U}\\    
          \hline                  
\end{tabular}
\end{table}

\subsection{Bilinear Groups}

Let $\mathbb{G}_{1}$, $\mathbb{G}_{2}$ and $\mathbb{G}_{\tau}$ be cyclic group with prime order $p$.  A map $e:\mathbb{G}_{1}\times\mathbb{G}_{2}\rightarrow\mathbb{G}_{\tau}$ is a bilinear group if the following properties are satisfied \cite{bf:ibe2001}:

\begin{enumerate}
\item{\sf Bilinearity.} For all $g\in\mathbb{G}_{1}$, $h\in\mathbb{G}_{2}$ and $x,y\in\mathbb{Z}_{p}$, $e(g^{x},h^{y})=e(g^{y},h^{x})=e(g,h)^{xy}$;
\item{\sf Non-degeneration.} For all $g\in\mathbb{G}_{1}$ and $h\in\mathbb{G}_{2}$, $e(g,h)\neq 1_{\tau}$ where $1_{\tau}$ is the identity element in $\mathbb{G}_{\tau}$;
\item{\sf Computability.} For all $g\in\mathbb{G}_{1}$ and $h\in\mathbb{G}_{2}$, there exists an efficient algorithm to compute $e(g,h)$.
\end{enumerate}
In the case that $\mathbb{G}_{1}=\mathbb{G}_{2}$, $e$ is called symmetric bilinear map. Let $\mathcal{BG}(1^{\ell})\rightarrow(e,p,\mathbb{G},\mathbb{G}_{\tau})$ be a  symmetric bilinear group generator which takes as input a security parameter $1^{\ell}$ and outputs a bilinear group $(e,p,\mathbb{G},\mathbb{G}_{\tau})$ with prime order $p$ and $e:\mathbb{G}\times\mathbb{G}\rightarrow\mathbb{G}_{\tau}$.

Note that Galbraith, Paterson and Smart \cite{gps:2008} classified parings into
three basic types and our scheme is based on the Type-I pairing where
$\mathbb{G}_{1}=\mathbb{G}_{2}$.
Our scheme uses these bilinear maps as required by the signatures schemes
described below.

\subsection{Complexity Assumptions}
\begin{definition}{\sf ($q$-Strong Diffie-Hellman (SDH) Assumption \cite{bb:ss2004})} Let $\mathcal{BG}(1^{\ell})\rightarrow(e,p,\mathbb{G},\mathbb{G}_{\tau})$, $g$ be a generator of $\mathbb{G}$ and $x\stackrel{R}{\leftarrow}\mathbb{Z}_{p}$. We say that $q$-strong Diffie-Hellman assumption holds on  $\mathbb{G}$ if for all probabilistic polynomial time (PPT) adversary $\mathcal{A}$ given $(g,g^{x},g^{x^{2}},\cdots,g^{x^{q}})$ can output a pair $(c,g^{\frac{1}{x+c}})$ with negligible probability, namely
$Adv_{\mathcal{A}}^{q-SDH}=\Pr\left[\mathcal{A}(g,g^{x},g^{x^{2}},\cdots,g^{x^{q}})\rightarrow(x,g^{\frac{1}{x+c}})\right]\leq \epsilon(\ell)$,
where $c\in\mathbb{Z}_{p}$.
\end{definition}

The security of the following two signatures used in our scheme and thus our
overall security can be shown to reduce to this complexity assumption.

\subsection{Zero-Knowledge Proof}
In this paper, we use zero-knowledge proof of knowledge protocols to prove knowledge of statements about discrete logarithms \cite{bg:pok2001}, including discrete logarithm, equality,  product, disjunction and conjunction. We follow the notation proposed in \cite{cs:gs1997} and formalised in \cite{cky:zop2009}. By $$\mbox{PoK}\left\{(\alpha,\beta,\gamma): A=g^{\alpha}h^{\beta}~\wedge~\tilde{A}=\tilde{g}^{\alpha}\tilde{h}^{\gamma}\right\},$$ we denote a zero-knowledge proof of knowledge of $\alpha,\beta$ and $\gamma$ such that $A=g^{\alpha}h^{\beta}$ and $\tilde{A}=\tilde{g}^{\alpha}\tilde{h}^{\gamma}$ holds in groups $\mathbb{G}$ and $\tilde{\mathbb{G}}$ simultaneously  where $\mathbb{G}=\langle g\rangle=\langle h \rangle$ and  $\tilde{\mathbb{G}}=\langle \tilde{g} \rangle=\langle \tilde{h} \rangle$.  Conventionally, the values in the parenthesis $(\alpha,\beta,\gamma)$ denote quantities of which knowledge is being proven, while the other values are public to the verifier. 

\subsection{Boneh-Boyen (BB) Signature}%
In 2004, Boneh and Boyen \cite{bb:ss2004} proposed a short signature scheme.
This scheme was used to construct efficient set-membership proof and range proof
\cite{ccs:sr2008}. In this paper, we use this signature scheme to generate tags
for the ticket policies. The scheme works as follows: \medskip

\noindent{\sf KeyGen.}  Let $\mathcal{BG}(1^{\ell})\rightarrow(e,p,\mathbb{G},\mathbb{G}_{\tau})$ and $g_{1},g_{2}$ be generators of $\mathbb{G}$. The signer generates a secret-public key pair $(x,Y)$ where $x\stackrel{R}{\leftarrow}\mathbb{Z}_{p}$ and $Y=g_{2}^{x}$.
\medskip

\noindent{\sf Signing.} To sign on a message $m\in \mathbb{Z}_{p}$, the signer computes the signature as $\sigma=g_{1}^{\frac{1}{x+m}}$. 
\medskip

\noindent{\sf Verifying.} To verify whether $\sigma$ is a signature on the message $m$, the verifier checks $e(\sigma,Yg_{2}^{m})\stackrel{?}{=}e(g_{1},g_{2})$. 
\medskip

\begin{theorem} (Boneh and Boyen \cite{bb:ss2004}) 
This signature is $(q_{S},\epsilon(\ell))$-secure against existentially forgery under the weak chosen message attacks if the $(q,\epsilon'(\ell))$-strong Diffie-Hellman (SDH) assumption holds on $(e,p,\mathbb{G},\mathbb{G}_{\tau})$, where $q_{S}$ is the number of signing queries made by the adversary $\mathcal{A}$,   $q>q_{S}$ and $\epsilon'(\ell)=\epsilon(\ell)$.
\end{theorem}

\subsection{Signature with Efficient Proof Protocol}%
Au {\em et al.} \cite{asm:ac2006} proposed a signature with an efficient proof
protocol scheme and referred to it as BBS+ signature. In this paper, we use
their signature scheme to issue credentials to users and ticket sellers and to
generate tickets for users. The scheme works as follows: \medskip

\noindent{\sf KeyGen.}  Let $\mathcal{BG}(1^{\ell})\rightarrow(e,p,\mathbb{G},\mathbb{G},\mathbb{G}_{\tau})$ and $(h,g_{0},g_{1},\cdots,$ $g_{n+1})$  be generators of $\mathbb{G}$.
The signer generates a secret-public key pair $(x,Y)$ where $x\stackrel{R}{\leftarrow}\mathbb{Z}_{p}$ and $Y=h^{x}$.
\medskip

\noindent{\sf Signing.} To sign on a block of messages $(m_{1},m_{2},\cdots,m_{n})\in\mathbb{Z}_{p}^{n}$, the signer selects $w,s\stackrel{R}{\leftarrow}\mathbb{Z}_{p}$ and computes $\sigma=(g_{0}g_{1}^{s}g_{2}^{m_{1}}\cdots g_{n+1}^{m_{n}})^{\frac{1}{x+w}}$. The signature on  $(m_{1},m_{2},\cdots,m_{n})$ is $(w,s,\sigma)$.
\medskip

\noindent{\sf Verifying.} To verify whether $(w,s,\sigma)$ is a valid signature on  $(m_{1},m_{2},$ $\cdots,m_{n})$, the verifier checks $e(\sigma,Yh^{w})\stackrel{?}{=}e(g_{0}g_{1}^{s}g_{2}^{m_{1}}\cdots g_{n+1}^{m_{n}},h)$. 
\medskip

\noindent{\em Proof of the Signature.} To prove $(w,s,\sigma)$ is a signature on
$(m_{1},m_{2},$ $\cdots,m_{n})$, the prover selects
$r_{1},r_{2}\stackrel{R}{\leftarrow}\mathbb{Z}_{p}$, and computes $A_{1}=\sigma
g_{2}^{r_{1}}$ and $A_{2}=g_{1}^{r_{1}}g_{2}^{r_{2}}$. Let $t_{1}=wr_{1}$ and
$t_{2}=wr_{2}$. %
In our scheme, we utilise Au {\em et al.}'s\cite{asm:ac2006} honest-verifier
zero-knowledge proof of knowledge protocol, $\Pi$, as follows:%
\begin{equation*}
\mbox{PoK}\left\{\begin{array}{ll}(r_{1},r_{2}, t_{1},t_{2},  w,s, \sigma,  m_{1},\cdots,m_{n}):  \\
A_{2}=g_{1}^{r_{1}}g_{2}^{r_{2}} \wedge A_{2}^{w}=g_{1}^{t_{1}}g_{2}^{t_{2}}\wedge \frac{e(A_{1},Y)}{e(g_{0},h)}=\\
e(g_{1},h)^{s}\cdot e(A_{1},Y)^{-w} \cdot e(g_{2},h)^{r_{1}w}\cdot\\
  e(g_{2},Y)^{r_{1}}\cdot \prod_{i=2}^{n+1}e(g_{i},h)^{m-1}
\end{array}\right\}.
\end{equation*}
\medskip

\begin{theorem} (Au {\em et al.} \cite{asm:ac2006} )
This signature with an efficient proof protocol is
$(q_{S},\epsilon(\ell))$-existentially unforgeable under the adaptively chosen
message attacks if the $(q,\epsilon'(\ell))$-strong Diffie-Hellman (SDH)
assumption holds on $(e,p,\mathbb{G},\mathbb{G}_{\tau})$, where $q_{S}$ is the
number of signing queries made by the adversary $\mathcal{A}$, $q>q_{S}$ and
$\epsilon(\ell)'>q\epsilon(\ell)$.
\end{theorem}

\section{Formal Definitions and Security Models}%
\label{def_security}%
In this section, we provide the formal definitions and security models of our
scheme which will be used to verify its security.

\subsection{Formal Definitions}\label{fd}

\begin{figure*}[!t]
\centering
\includegraphics[height=6.5cm ,width=13cm]{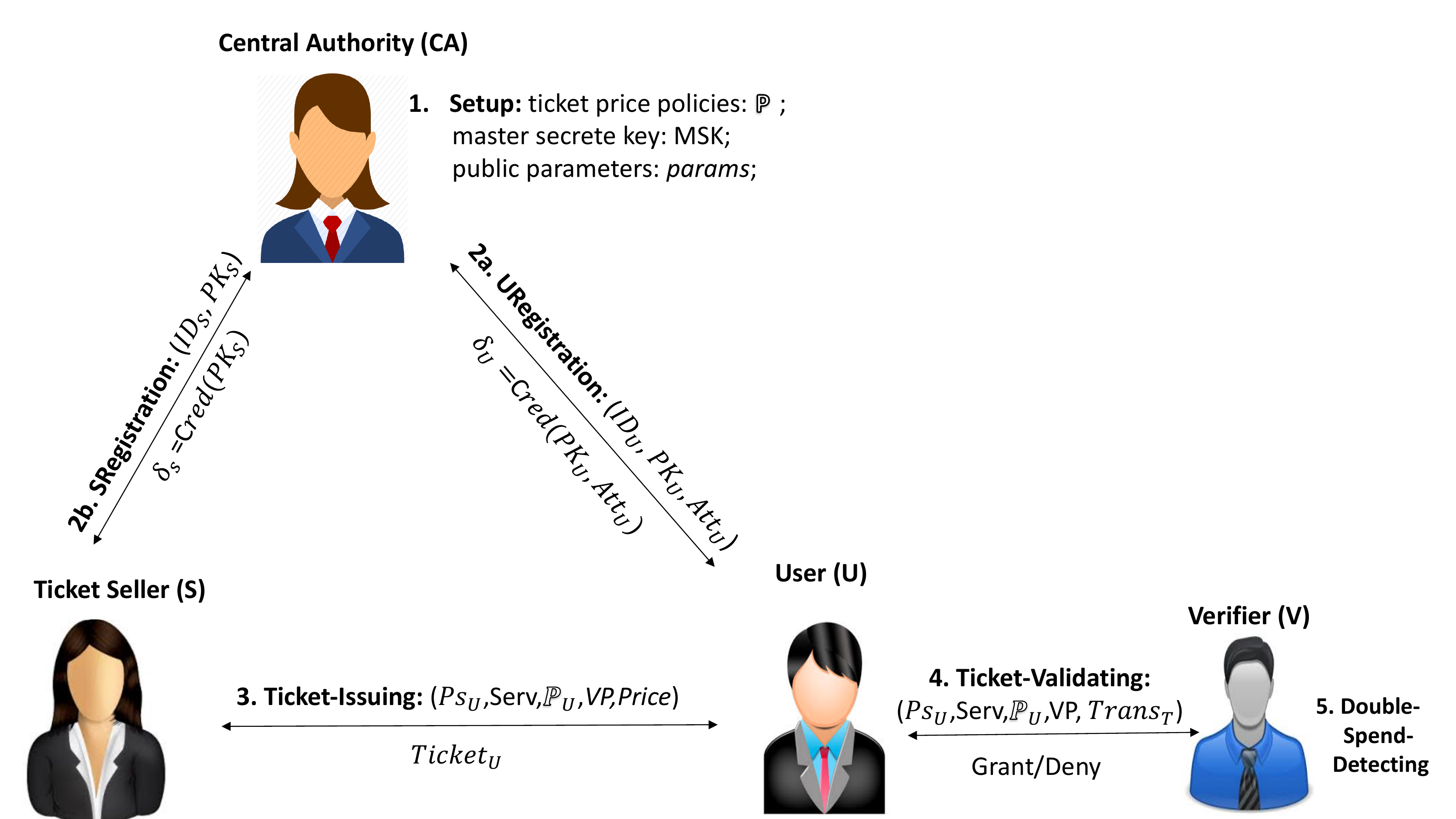}\caption{The Model 
of our scheme}\label{model}
\end{figure*}

Our scheme consists of the following four entities: central authority {\sf CA}, 
user {\sf U}, ticket seller {\sf S} and ticket verifier {\sf V}.
\begin{itemize}
\item{\sf CA} authenticates {\sf U} and {\sf S}, and issues anonymous
credentials to them; \item{\sf S} registers to the  {\sf CA}, obtains anonymous
credentials from the {\sf CA}, and sells tickets to {\sf U} in accordance with
the ticket policies; \item {\sf U} registers to the {\sf CA}, obtains anonymous
credentials from the {\sf CA}, purchases tickets from {\sf S}, and proves the
possession of tickets to {\sf V}; \item{\sf V} validates the tickets provided by
{\sf U} and detects whether a ticket is double spent.
\end{itemize}

The interactions between the different entities in our scheme is shown in Fig.
\ref{model}. The algortihms associated with these interactions are formally
defined as follows: \medskip

\noindent{\sf Setup}$(1^{\ell})\rightarrow(MSK,params,\mathbb{P})$.  {\sf CA} inputs  $1^{\ell}$, and outputs  $MSK$, $params$ and    $\mathbb{P}$.
\medskip

\noindent{\sf Registration}. This algorithm consists of the following two sub-algorithms: {\sf S}'s registration {\sf SRegistration} and {\sf U}'s registration {\sf URegistration}.

\begin{enumerate}
\item{\sf SRegistration}$({\sf S}(ID_{S},SK_{S},PK_{S},params)\leftrightarrow {\sf CA}(MSK,PK_{S},$ $params))\rightarrow(\sigma_{S}, (ID_{S},PK_{S}))$. {\sf S} runs $\mathcal{KG}(1^{\ell})\rightarrow(SK_{S},$ $PK_{S})$ to generate  $(SK_{S},PK_{S})$, inputs  $ID_{S}$,  $(SK_{S},$ $PK_{S})$ and  $params$, and outputs $\sigma_{S}$ which is generated by {\sf CA}. {\sf CA} inputs $MSK$,  $PK_{S}$ and  $params$, and outputs $(ID_{S},PK_{S})$.
\medskip
\item{\sf URegistration}$({\sf U}(ID_{U},A_{U}, SK_{U},PK_{U},params)$ $\leftrightarrow{\sf CA}(MSK,A_{U}, PK_{U},params))\rightarrow (\sigma_{U},(ID_{U},$ $PK_{U}))$. {\sf U} runs $\mathcal{KG}(1^{\ell})\rightarrow(SK_{U},PK_{U})$ to generate $(SK_{U},PK_{U})$, inputs  $ID_{U}$,   $A_{U}$,  $(SK_{U},PK_{U})$ and  $params$,  and outputs $\sigma_{U}$ which is generated by {\sf CA}. {\sf CA} inputs  $MSK$,  $A_{U}$,  $PK_{U}$ and  $params$, and outputs  $(ID_{U},PK_{U})$.
\end{enumerate}

\noindent{\sf Ticket-Issuing}$({\sf U}(SK_{U},PK_{U},A_{U},\sigma_{U},Ps_{U},\mathbb{P},VP,Serv,$ $params)\leftrightarrow {\sf S}(SK_{S},PK_{S},Ps_{U},\mathbb{P},Price,VP,Serv,$ $params))\rightarrow(Ticket_{U},(Ps_{U},Service))$. This is an interactive algorithm executed between {\sf U} and {\sf S}. {\sf U} inputs  $(SK_{U},PK_{U})$,   $A_{U}$,   $\sigma_{U}$,  $Ps_{U}$,  $\mathbb{P}$, $VP$,  $Serv$ and  $params$, and outputs   $Ticket_{U}$. {\sf S} inputs $(SK_{S},PK_{S})$,  $Ps_{U}$,  $\mathbb{P}$,  $Price$,  $VP$,  $Serv$ and  $params$, and outputs $(Ps_{U},Serv)$.

\noindent{\sf Ticket-Validating}$({\sf
U}(SK_{U},Ps_{U},Ticket_{U},VP,Serv,params)$ $\leftrightarrow {\sf V
}(VP,Serv,params))\rightarrow(0/1,(Serv, $ $Trans_{T}))$. This is an interactive
algorithm executed between {\sf U} and {\sf V}. {\sf U} inputs
$(SK_{U},PK_{U})$,     $Ticket_{U}$,  $VP$,  $Serv$ and  $params$, and outputs 1
if  $Ticket_{U}$ is valid; otherwise it outputs $0$ to indicate a failure. {\sf
V} inputs  $VP$, $Serv$ and  $params$, and outputs $(Serv,Trans_{T})$.


\noindent{\sf
Double-Spend-Detecting}$(Trans_{T},params)\rightarrow(PK_{U},\perp)$. {\sf V}
inputs $Trans_{T}$ and  $params$ and outputs $PK_{U}$ if {\sf U} has used a 
ticket twice; otherwise it outputs $\perp$.

\begin{definition}
Our scheme is correct if 

\begin{equation*}
\Pr\left[\begin{array}{l|l} 
  & {\sf Setup}(1^{\ell})\rightarrow(msk,params,\mathbb{P});\\
  {\sf Ticket-}    & {\sf SRegistration}({\sf S}(ID_{S},SK_{S}, \\
    {\sf Validating}            &           PK_{S}, params)\leftrightarrow {\sf CA}(MSK,\\
     ( {\sf U}(SK_{U},                        &      PK_{S},    params)) \rightarrow(\sigma_{S},(ID_{S},\\
         Ps_{U},                                     & PK_{S}));\\
      Ticket_{U},    &        {\sf URegistration}({\sf U}(ID_{U},A_{U}, \\
    VP, Serv,                 &        SK_{U}, PK_{U},params)\leftrightarrow \\
   params)  &       {\sf CA}(MSK, PK_{U},  A_{U},params)) \\
  \leftrightarrow{\sf V}(VP,                                & \rightarrow(\sigma_{U}, (ID_{U},PK_{U}));\\
   Serv,               &         {\sf Ticket-Issuing}({\sf U}(SK_{U},PK_{U},\\
    params))                           &     A_{U},   \sigma_{U}, Ps_{U},\mathbb{P},VP,Serv,\\
 \rightarrow(1,   & params)\leftrightarrow{\sf S}(SK_{S},PK_{S},\\
(Serv,  &Ps_{U},\mathbb{P},VP,Serv, params))\\
Trans_{T}))& \rightarrow(Ticket_{U},(Ps_{U},Serv));\\
& A_{U}\models \mathbb{P}\\
\end{array}\right]=1
\end{equation*}
and 

\begin{equation*}
\Pr\left[\begin{array}{l|l} & {\sf Setup}(1^{\ell})\rightarrow(msk,params,\mathbb{P});\\
& {\sf SRegistration}({\sf S}(ID_{S},SK_{S}, \\
 & PK_{S},params)\leftrightarrow {\sf CA}(MSK,\\
& PK_{S},params))\rightarrow(\sigma_{S},(ID_{S},\\
& PK_{S}));\\
    {\sf Double-}                  &             {\sf URegistration}({\sf U}(ID_{U},A_{U}, \\
    {\sf Spend-}              &        SK_{U},  PK_{U},params)\leftrightarrow  \\
  {\sf Detecting}                        &    {\sf CA}(MSK,   PK_{U},  A_{U},params)) \\
 (Trans_{T},                                     &     \rightarrow   (\sigma_{U}, (ID_{U},PK_{U}));\\
params)                        &         {\sf Ticket-Issuing}({\sf U}(SK_{U},PK_{U},\\
\rightarrow PK_{U}      &  A_{U}, \sigma_{U},Ps_{U},\mathbb{P},VP,Serv,\\
&params)\leftrightarrow{\sf S}(SK_{S},PK_{S},Ps_{U},\\
& \mathbb{P},VP,Service,params))\rightarrow\\
&(T_{U},(Ps_{U},Service));\\
& A_{U}\models \mathbb{P}~ \wedge ~T_{U} ~\mbox{is double spent}.\\
\end{array}\right]=1.
\end{equation*}

\end{definition}

\subsection{Security Model}%
While Universally Composable  (UC) security models \cite{can:uc2001} can offer
strong security, it is very difficult to construct a scheme which can be shown
to provide UC security. To the best of our knowledge, none of the existing smart
ticketing schemes was proven in UC security model. Consequently, the security of
our scheme is defined by using the simulation-based definition as introduced in
\cite{cns:2007,gh:2007,cdn:2009,ci:2017}. The simulation-based model is defined
by the indistinguishability between the following ``real world'' and ``ideal
world'' experiment. \medskip

\noindent{\em The Real-World Experiment.}  We first present how our scheme works
where the central authority {\sf CA}, the ticket seller {\sf S}, the user {\sf
U} and the ticket verifier {\sf V} are honest. The real-world adversary
$\mathcal{A}$ can control  {\sf S}, {\sf U} and  {\sf V}, but cannot control
{\sf CA}. The entities controlled by $\mathcal{A}$ can deviate arbitrarily from
their behaviour described below.
{\sf CA} runs {\sf Setup}$(1^{\ell})\rightarrow(MSK,params,\mathbb{P})$ to 
generate the master secret key $msk$, system public parameters $params$ and the 
universal set $\mathbb{P}$ of ticket polices, and sends  $params$ and  
$\mathbb{P}$ to {\sf U}, {\sf S} and {\sf V}.

When receiving a registration message $(registration,ID_{S})$ from 
$\mathcal{E}$, {\sf S} executes the  seller registration algorithm {\sf 
SRegistration} with {\sf CA}. {\sf S} runs 
$\mathcal{KG}(1^{\ell})\rightarrow(SK_{S},PK_{S})$, takes as input his identity 
$ID_{S}$, the secret-public key pair $(SK_{S},PK_{S})$ and the public 
parameters $params$, outputs a credential $\sigma_{S}$. {\sf CA} takes inputs 
his master secret key $MSK$, {\sf S}'s public key $PK_{S}$ and the public 
parameters $params$, and outputs {\sf S}'s identity $ID_{S}$ and public key 
$PK_{S}$. {\sf S} sends a bit $b\in\{0,1\}$ to $\mathcal{E}$ to show  whether 
the {\sf SRegistation} algorithm succeed $(b=1)$ or failed $(b=0)$.

When receiving a registration message $(registration,ID_{U},A_{U})$ from $\mathcal{E}$, {\sf U} executes the  user registration algorithm {\sf URegistration} with {\sf CA}. {\sf U} runs $\mathcal{KG}(1^{\ell})$ $\rightarrow(SK_{U},PK_{U})$, takes as input his identity $ID_{U}$, attributes $A_{U}$,  secret-public key pair $(SK_{U},PK_{U})$ and the public parameters $params$, and outputs a credential $\sigma_{U}$. {\sf CA} takes inputs his master secret key $MSK$, {\sf U}'s public key $PK_{U}$ and the public parameters $params$, and outputs {\sf U}'s identity $ID_{U}$, attributes $A_{U}$ and public key $PK_{U}$. {\sf U} sends a bit $\tilde{b}\in\{0,1\}$ to $\mathcal{E}$ to show  whether the {\sf URegistation} algorithm succeed $(\tilde{b}=1)$ or failed $(\tilde{b}=0)$.

When receiving a ticket issuing message $(ticket$ $\_issuing,A_{U},VP,Service)$ 
from $\mathcal{E}$, {\sf U} first checks whether he has got a credential for 
$A_{U}$. If so, {\sf U} executes the ticket issuing algorithm {\sf 
Ticket-Issuing} with {\sf S}. {\sf U} takes as inputs his secret-public key 
pair $(SK_{U},PK_{U})$, attributes $A_{U}$,  a pseudonym $Ps_{U}$, his 
credential $\sigma_{U}$, the valid period $VP$, the service $Serv$ and the 
public parameters $params$. {\sf S} takes  as input his secret-public key pair 
$(SK_{S},PK_{S})$, the valid period $VP$, the service $Serv$ and the public 
parameters $params$.   Finally, {\sf U} obtains a ticket $T_{U}$ or $\perp$ to 
show failure. {\sf S}  outputs {\sf U}'s pseudonym $Ps_{U}$ and the service 
$Serv$. If the ticket issue is successful, {\sf U} sends a bit 
$\check{b}\in\{0,1\}$ to $\mathcal{E}$ to show the {\sf Ticket-Issuing} 
algorithm succeed $(\check{b}=1)$ or failed $(\check{b}=0)$.

When receiving a ticket validation message $(ticket\_validating,T_{U}, VP, Serv, params)$ from $\mathcal{E}$, {\sf U} first checks whether he has the ticket $T_{U}$ which includes the valid period $VP$ and the service $Serv$. If so, {\sf U} executes the ticket validating algorithm {\sf Ticket-Validating} with {\sf V}; otherwise {\sf U} outputs $\perp$ to show he does not have the ticket $T_{U}$. If {\sf U} has the ticket $T_{U}$, he takes as input his secret-public key pair $(SK_{U},PK_{U})$, the ticket $T_{U}$, the valid period  $VP$, the service $Serv$ and the system public parameters $params$, and outputs a bit $\hat{b}\in\{0,1\}$ to show whether the ticket is valid $(\hat{b}=1)$ or invalid $(\hat{b}=0)$. {\sf V} takes input the valid period $VP$, the service $Serv$ and the public parameters $params$, and outputs the service $Serv$ and the transcript $Trans$. Finally, if $\hat{b}=1$, {\sf U} returns $success$; otherwise {\sf U} returns $fail$.

When receiving a double spend detecting message 
$(double\_spend\_detecting,Trans,params)$ from $\mathcal{E}$, {\sf V} checks 
that whether there is a $(Trans',params)$ with $Trans=Trans'$. If so, {\sf V} 
returns a bit $\bar{b}=1$ to indicate that it is a double spend ticket; 
otherwise $\bar{b}=0$ is returned to show that the ticket has not been double 
spent. 
\vspace{0.05cm}

\noindent{\em The Ideal-World Experiment.} %
In the ideal world experiment, there are the same entitles as in real world
experiment, including the  central authority  ${\sf CA}'$, ticket seller ${\sf
S}'$, user ${\sf U}'$ and ticket verifier ${\sf V}'$. All communications among
these entities must go through a trusted party {\sf TP}. The behaviour of {\sf
TP} is described as follows. {\sf TP} maintains four lists which are initially
empty: a ticket seller credential list, a user credential list, a
ticket list for each user and a ticket validating list.

When receiving a registration message $(registration,ID_{S'})$ from ${\sf S}'$,
{\sf TP} sends $( registration, $ $ID_{S'})$ to ${\sf CA}'$ and obtains a bit
$\nu\in\{0,1\}$ from ${\sf CA}'$. If $\nu=1$, {\sf TP} adds ${\sf S}'$ into the
ticket seller credential list and sends $\nu$ to ${\sf S}'$; otherwise, {\sf TP}
sends $\nu=0$ to ${\sf S}'$ to indicate failure.

When receiving a registration message $(registration,ID_{U'},A_{U'})$ from 
${\sf U}'$, {\sf TP} sends $(registration,ID_{U'},A_{U'})$ to ${\sf CA}'$ and 
obtains a bit $\tilde{\nu}\in\{0,1\}$ from ${\sf CA}'$. If $\tilde{\nu}=1$, 
{\sf TP} adds $({\sf U'},A_{U'})$ into the user credential list and sends 
$\tilde{\nu}$ to ${\sf U}'$; otherwise, ${\sf TP}$ sends $\tilde{\nu}=0$ to 
${\sf S}'$ to indicate failure.

When receiving a ticket issuing message 
$(ticket\_issuing,Ps_{U},Price,VP,Serv)$ from ${\sf U}'$, {\sf TP} sends 
$(ticket\_issuing,Ps_{U},Price,VP,Serv)$ to ${\sf S}'$ and obtains a bit 
$\hat{\nu}\in\{0,1\}$ from ${\sf S}'$. If $\hat{\nu}=1$, {\sf TP} adds $({\sf 
U'},A_{U'},Ps_{U},Price,VP,Serv)$  into the user ticket list, and sends 
$\hat{\nu}$ to ${\sf V}'$; otherwise, {\sf TP} sends $\hat{\nu}=0$ to ${\sf 
U}'$ to indicate failure.

When receiving a ticket validating message $(ticket\_$ $validating,$ $T_{U'})$ 
from ${\sf V}'$, {\sf TP} checks whether $T_{U'}$ is in the user ticket list. 
If so,  {\sf TP}  sends a bit $\bar{\nu}=1$  to ${\sf U}'$ and puts $T_{U'}$ 
into $UVL$; otherwise, ${\sf TP}'$ sends $\bar{\nu}=0$ to indicate  failure.

When receiving a double spend detecting message $(double\_$ $spend\_detecting, 
T_{U'})$ from ${\sf V}'$, {\sf TP} checks whether $T_{U}\in UVL$. If it is, 
{\sf TP} returns $\check{\nu}=1$ to ${\sf U}'$ to indicate it is double spend; 
otherwise, $\check{\nu}=0$ is returned to show it is not double spent.

The entities ${\sf CA}'$, ${\sf S}'$, ${\sf U}'$ and ${\sf V}'$ in ideal world simply relay the inputs and outputs between $\mathcal{E}$ and {\sf TP}.

\begin{definition}%
Let {\bf Real}$_{\mathcal{E},\mathcal{A}}(\ell)$ be the probability that the
environment $\mathcal{E}$ outputs $1$ when running in the real world with the
adversary $\mathcal{A}$ and {\bf Ideal}$_{\mathcal{E},\mathcal{A}'}$ be the
probability that $\mathcal{E}$ outputs 1 when running in the ideal world with
the adversary $\mathcal{A}'$. A set of cryptographic protocols is said to
securely implement our scheme if
$ \left|{\bf Real}_{\mathcal{E},\mathcal{A}}(\ell)-{\bf Ideal}_{\mathcal{E},\mathcal{A}'}(\ell)\right|\leq \epsilon(\ell).$
 \end{definition}
\medskip

\noindent{\bf Security Properties.} We now look at the security properties of
our scheme which the ideal-world experiments can provide.%
\medskip

\noindent{\em User's Privacy.} %
${\sf S}'$ does not know users' identities and their exact attributes, namely
${\sf S}'$ only knows that a user buys a ticket for which  she has the required
attributes. Even if ${\sf S}'$ colludes with ${\sf V}'$ and potentially with
other users, they can only try to know the attributes required by the ticket
policies. Furthermore, two tickets for the same users cannot be linked. Since
each user needs to prove that he knows the corresponding secret key included in
a ticket when using the ticket, he cannot transfer his tickets to others.
untransferability.

\noindent{\em Seller's Security.}%
 ${\sf U}'$ cannot generate a ticket on behalf of the seller ${\sf S}'$. Even if
${\sf U}'$ colludes potentially with other users and ${\sf V}'$, they cannot
forge a valid ticket. Since a double spend ticket can be detected and the real
user can be identified, ${\sf U'}$ cannot double spend a ticket. Therefore, the
seller's security includes both unforgeability, double spend detection and
de-anonymization. \medskip

In Section \ref{secu}, we prove the indistinguishability between the real-world
experiments and ideal-world experiments and hence show that the above security
properties can be achieved.

\section{Construction of our scheme}%
\label{const}%
In this section, we describe the formal construction of our scheme. Our scheme
uses a number of ideas and concepts from Au\etal's signature with efficient
protocol scheme \cite{asm:ac2006}, Camenisch\etal's set-membership proof scheme
and range proof scheme \cite{ccs:sr2008}, Pedersen's commitment scheme
\cite{p:com1991} and Au\etal's e-cash \cite{asm:ec2008} scheme. In particular,
we incorporated Au\etal's signature scheme which enables a user to obtain a
signature on a committed block of attributes and prove the knowledge of the
signature in zero-knowledge. This is  to issue credentials to users and ticket
sellers and to generate tickets for users.  We adapt
Camenisch\etal\cite{ccs:sr2008}'s set-membership proof and range proofs schemes
to prove a user's attributes. In these schemes, a user can prove to the verifier
that an attribute is in a set or in a range without the verifier knowing the
exact value. In our scheme, these attributes are additionally certified by a
trusted third party as well. Moreover, multiple sets and ranges were not
considered simultaneously, whereas our scheme does. Pedersen's commitment scheme
is used in our scheme to hide the knowledge which a prover needs to prove. And
lastly we incorporate Au\etal's \cite{asm:ec2008} approach to detect and
de-anonymise a double spend user.

\medskip

\noindent{\em Construction challenges:} %
The schemes described in \cite{asm:ac2006}, \cite{asm:ec2008},
\cite{ccs:sr2008}, \cite{p:com1991}  form the basis of our construction, the
challenge is to combine and adapt them such that the resulting scheme provides
the following three additional features: (1) The attributes (\eg age,
disability, \etc) which a user needs to prove to a ticket seller must be
certified by a trusted third party or otherwise users could simply buy
discounted tickets using attributes which they do not possess. To address this,
Au\etal's signature scheme \cite{asm:ac2006} is used to certify a user's
attributes. As a result, all the values which are included in the credentials
are expressed as discrete logarithm formulas which can then be proven using the
zero-knowledge proof of knowledge protocol proposed by Bellare\etal in
\cite{bg:pok2001}. (2) Tickets need to be untransferable and unlinkable while
doublespend detection must be possible. Thus our tickets are generated using
anonymous credentials (unlinkability) which include a user's personal
information (untransferability). To detect a doublespend user, each ticket is
includes a serial number. If two tickets have the same serial number, the public
trace technique proposed by Au\etal in \cite{asm:ec2008} is used to reveal the
user's identity (via her public key). (3) To provide a high degree of
flexibility for setting ticket policies, Camenisch\etal's range and
set-membership proofs \cite{ccs:sr2008} must both be available for use
simultaneously. User can then use their certified attributes to demonstrate
membership of multiple range and set policies, \eg to get a young-persons
discount, a frequent traveller bonus as well as a disability reduction.

\subsection{High-Level Overview} In our e-ticket system, the type of ticket can
be influenced by two kinds of policies: range  and set. Range policies might
include attributes like age, number of journeys made, salary, \etc ; while set
policies might consist of various other attributes, such as profession,
disability, location, \etc%
~Our scheme allows users to anonymously prove their attributes to a ticket
seller and works as follows:\\

\noindent{\em Setup.} %
Figure~\ref{setup} shows how the scheme is initialised. The ticket price
polices, $\mathbb{P}$, is set to $\mathbb{P}= \big\{\mathbb{R}_{1},
\cdots,\mathbb{R}_{N_{1}}, \mathbb{S}_{1}, \cdots,\mathbb{S}_{N_{2}}\big\}$
where $\{\mathbb{R}_{1}, \cdots, \mathbb{R}_{N_{1}}\}$ are the supported range
policies and $\{\mathbb{S}_{1}, \cdots, \mathbb{S}_{N_{2}}\}$ are the supported
set policies. 
The {\sf CA} selects the following secret keys
$MSK=(x,y,\mu_{1},\mu_{2},\cdots,\mu_{N_{2}})$ where $x$ is used to generate
credentials for users of the system, $y$ is used to generate tags identifying
the range policies and the $\mu_{i}$s ($i=1,2,\cdots,N_{2}$) are used to
generate tags identifying the set policies. The {\sf CA} then publishes it
public parameters, $params$, which include the ticket price policy,
$\mathbb{P}$, together with the range and %
and set policy tags %
as well as a number of other values required by the scheme. 
%
\medskip

\noindent{\em Registration.} The steps involved in the registration process are
shown in Figure~\ref{regist}. The registration of a seller, {\sf S}, requires
{\sf S} to generate a secret-public key pair $(x_{s},Y_{S})$. He sends $Y_S$ to
the {\sf CA} as well as a proof of knowledge, $\Pi_S^1$, to demonstrate he knows
the secret key $x_s$. Using some out-of-band channel, {\sf S} authenticates
himself to the {\sf CA} and provides evidence that he is allowed to operate as a
seller. If $PI_S^1$ is valid and the authentication is successful, the {\sf CA}
computes a credential, $\sigma_{S}$ as part of a BBS+ signature scheme which
includes the public key $Y_S$ as well as a validity period for it, $VP_S$. These
details are then sent back to {\sf S} who uses his private and public keys, the
validity period, $VP_S$, and their associated BBS+ signature to verify that the 
{\sf CA} has authorized him as a seller.

In the case of a user registration, a user {\sf U}  generates a secret-public
key pair $(x_{u},Y_{U})$ and submits her public key together with her a proof of
knowledge, $\Pi_U^1$ showing that she knows the secret key, $x_{u}$. She also
sends the {\sf CA} the list of attributes, $A_{U}$, (\eg age,  profession,
location, {\em etc.}) which allow her to get discounted tickets. Again, using an
out-of-band channel, she authenticates herself to the {\sf CA} and provides
evidence for the claimed attributes. If $\prod_U^1$ holds, the authentication is
successful and the {\sf CA} is satisfied with the provided evidence,  it
computes a credential $\sigma_U$ as part of a BBS+ signature scheme which
includes the public key $Y_U$, its validity period $VP_U$ as well as the
corresponding range and set tags of the user's attributes. %
These details are sent back to {\sf U} who uses them to verify that she is now a
legitimate user of the system and that her attributes have been certified by the
{\sf CA}. \vspace{0.05cm}

\begin{figure*}[!h]
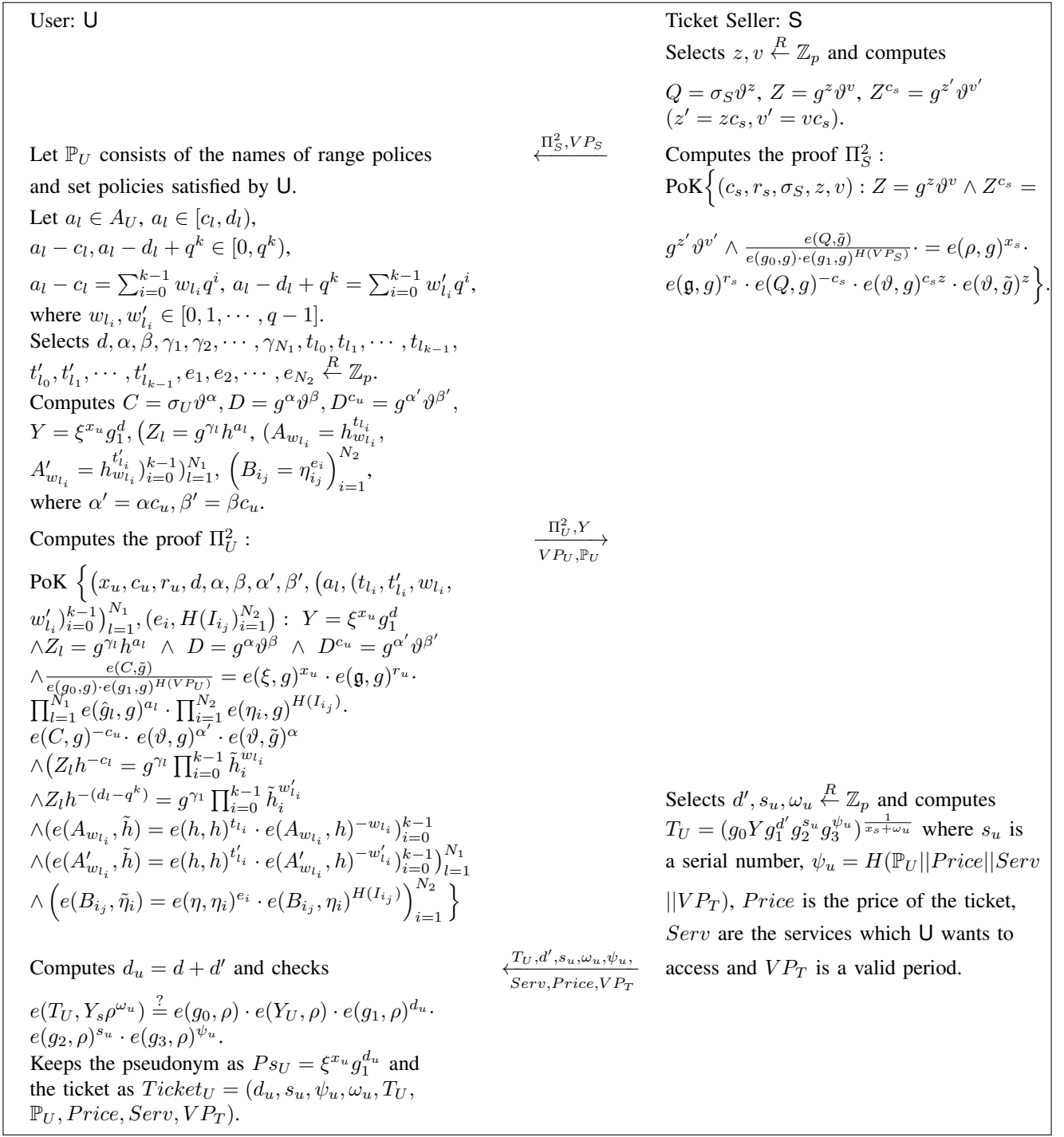

\centering
\fbox{
\begin{minipage}{17cm}

\begin{tabular}{lcl}
User: {\sf U} & & Ticket Seller: {\sf S}\\
\medskip

& & Selects $z,v\stackrel{R}{\leftarrow}\mathbb{Z}_{p}$ and computes \\
& & $Q=\sigma_{S}\vartheta^{z}$, $Z=g^{z}\vartheta^{v}$, $Z^{c_{s}}=g^{z'}\vartheta^{v'}$\\ 
& &  $(z'=zc_{s}, v'=vc_{s})$.\\
Let  $\mathbb{P}_{U}$ consists of the  names of  range polices &   $\xleftarrow{\Pi_{S}^{2},VP_{S}}$ & Computes the proof $\Pi_{S}^{2}:$\\
and set  policies satisfied by {\sf U}. & & $\mbox{PoK}\Big\{ (c_{s},r_{s},\sigma_{S},z,v):
 Z=g^{z}\vartheta^{v}\wedge Z^{c_{s}}=$\\
Let $a_{l}\in A_{U}$, $a_{l}\in[c_{l},d_{l})$, \\
 $a_{l}-c_{l},a_{l}-d_{l}+q^{k}\in[0,q^{k})$, & & $g^{z'}\vartheta^{v'} \wedge
\frac{e(Q,\tilde{g})}{e(g_{0},g)\cdot e(g_{1},g)^{H(VP_{S})}}\cdot =e(\rho,g)^{x_{s}}\cdot $\\
 $a_{l}-c_{l}=\sum_{i=0}^{k-1}w_{l_{i}}q^{i}$,    $a_{l}-d_{l}+q^{k}=\sum_{i=0}^{k-1}w'_{l_{i}}q^{i}$,  && $ e(\mathfrak{g},g)^{r_{s}}
\cdot e(Q,g)^{-c_{s}}\cdot  e(\vartheta,g)^{c_{s}z}\cdot e(\vartheta,\tilde{g})^{z}\Big\}.$\\
  where $w_{l_{i}},w_{l_{i}}'\in[0,1,\cdots,q-1]$.  && \\

Selects $d,\alpha,\beta,\gamma_{1},\gamma_{2},\cdots,\gamma_{N_{1}},t_{l_{0}},t_{l_{1}},\cdots,t_{l_{k-1}},$&& 
 \\
$t'_{l_{0}},t'_{l_{1}},\cdots,t'_{l_{k-1}},e_{1},e_{2},\cdots,e_{N_{2}}\stackrel{R}{\leftarrow}\mathbb{Z}_{p}.$&&\\  
Computes
$C=\sigma_{U}\vartheta^{\alpha},D=g^{\alpha}\vartheta^{\beta}, D^{c_{u}}=g^{\alpha'}\vartheta^{\beta'},$ & & \\

$Y=\xi^{x_{u}}g_{1}^{d},\big(Z_{l}=g^{\gamma_{l}}h^{a_{l}}$,
$(A_{w_{l_{i}}}=h_{w_{l_{i}}}^{t_{l_{i}}},$ & & \\$A'_{w_{l_{i}}}=h_{w_{l_{i}}}^{t'_{l_{i}}})_{i=0}^{k-1})_{l=1}^{N_{1}},$
$\left(B_{i_{j}}=\eta_{i_{j}}^{e_{i}}\right)_{i=1}^{N_{2}}$, \\
where $\alpha'=\alpha c_{u},\beta'=\beta c_{u}$.\\

Computes the proof $\Pi_{U}^{2}:$ & $\xrightarrow[VP_{U},\mathbb{P}_{U}]{\Pi_{U}^{2},Y}$ &\\
\mbox{PoK}
$\Big\{
\big(x_{u}, c_{u},r_{u}, d,\alpha,\beta,\alpha',\beta', \big(a_{l}, (t_{l_{i}},t'_{l_{i}},w_{l_{i}},$\\
$w'_{l_{i}})_{i=0}^{k-1}\big)_{l=1}^{N_{1}},  (e_{i}, H(I_{i_{j}})_{i=1}^{N_{2}}\big):~Y=\xi^{x_{u}}g_{1}^{d}$\\
$ \wedge Z_{l}=g^{\gamma_{l}}h^{a_{l}} ~\wedge~ D=g^{\alpha}\vartheta^{\beta}~\wedge~D^{c_{u}}=g^{\alpha'}\vartheta^{\beta'} $\\
$\wedge \frac{e(C,\tilde{g})}{e(g_{0},g)\cdot e(g_{1},g)^{H(VP_{U})}}=e(\xi,g)^{x_{u}}\cdot e(\mathfrak{g},g)^{r_{u}}\cdot $ \\
 $\prod_{l=1} ^{N_{1}}e(\hat{g}_{l},g)^{a_{l}}\cdot   \prod_{i=1}^{N_{2}}e(\eta_{i},g)^{H(I_{i_{j}})} \cdot$\\
  $e(C,g)^{-c_{u}}\cdot$ 
  $   e(\vartheta,g)^{\alpha'}\cdot    e(\vartheta,\tilde {g})^{\alpha}$ \\
 $ \wedge \big(Z_{l}h^{-c_{l}}=g^{\gamma_{l}}\prod_{i=0}^{k-1}\tilde{h}_{i}^{w_{l_{i}}} $\\
$ \wedge Z_{l}h^{-(d_{l}-q^{k})}=g^{\gamma_{1}}\prod_{i=0}^{k-1}\tilde{h}_{i}^{w'_{l_{i}}}$ & & Selects $d',s_{u},\omega_{u}\stackrel{R}{\leftarrow}\mathbb{Z}_{p}$ and computes\\
 $ \wedge
  (e(A_{w_{l_{i}}},\tilde{h})=e(h,h)^{t_{l_{i}}}\cdot e(A_{w_{l_{i}}},h)^{-w_{l_{i}}})_{i=0}^{k-1}$ & & $T_{U}=(g_{0}Yg_{1}^{d'}g_{2}^{s_{u}}g_{3}^{\psi_{u}})^{\frac{1}{x_{s}+\omega_{u}}}$  where  $s_{u}$ is  \\
  $\wedge
    (e(A'_{w_{l_{i}}},\tilde{h})=e(h,h)^{t'_{l_{i}}}\cdot e(A'_{w_{l_{i}}},h)^{-w'_{l_{i}}})_{i=0}^{k-1}\big)_{l=1}^{N_{1}}$ & & a serial number, $\psi_{u}=H(\mathbb{P}_{U}||Price||Serv$\\
     $\wedge \left(e(B_{i_{j}},\tilde{\eta}_{i})=e(\eta,\eta_{i})^{e_{i}}\cdot e(B_{i_{j}},\eta_{i})^{H(I_{i_{j}})}\right)_{i=1}^{N_{2}}
\Big\}$ & &  $||VP_{T})$, $Price$ is the price of  the ticket,  \\
& & $Serv$ are the services  which  {\sf U}   wants  to  \\ 
Computes $d_{u}=d+d'$ and checks & $\xleftarrow[Serv,Price,VP_{T}]{T_{U},d',s_{u},\omega_{u},\psi_{u},}$ & access and $VP_{T}$ is a valid period. \\
 $e(T_{U},Y_{s}\rho^{\omega_{u}})\stackrel{?}{=}e(g_{0},\rho)\cdot e(Y_{U},\rho)\cdot e(g_{1},\rho)^{d_{u}}\cdot$ &  &    \\

 $ e(g_{2},\rho)^{s_{u}}\cdot e(g_{3},\rho)^{\psi_{u}}$.& & \\

Keeps the pseudonym as  $Ps_{U}=\xi^{x_{u}}g_{1}^{d_{u}}$  and \\
the ticket as  $Ticket_{U}=(d_{u},s_{u},\psi_{u},\omega_{u},T_{U},$& &\\
$\mathbb{P}_{U},Price,Serv,VP_{T})$.
\end{tabular}
\end{minipage}
}
\caption{Ticket Issuing Algorithm} \label{issue}
\end{figure*}

\noindent{\em Ticket Issuing.} Figure~\ref{issue} shows the details of the
ticket issuing phase. In order to prevent attackers from collecting users'
private information, a seller {\sf S} first needs to prove to {\sf U} that he is
authorised by the {\sf CA}. This is done by constructing a proof of knowledge,
$\Pi_{S}^{2}$ involving the seller's credential $\sigma_S$. If the proof holds,
the user {\sf U} proceeds by generating a new pseudonym, $Y$, which involves her
private key $x_{u}$ and constructs a proof of knowledge, $\Pi_{U}^{2}$. This
proof shows to {\sf S} that the {\sf CA} has certified her as a legitimate user
who has the claimed attributes which entitle her to buy the ticket corresponding
to her provided attributes. After {\sf S} has successfully verified her proof,
he constructs $T_{U}$ applying a BBS+ signature scheme which includes the user's
pseudonym, $Y$, the applicable range and set policies of the user relevant to
the ticket, a serial number to enable double spend detection as well as the
ticket's price and validity period, $VP_{T}$. Note that while the ticket price
and its validity period are included in the construction of $T_U$, they are just
free text entries and should only be used when price and validity periods are
required by the application context, \eg  when the validity period is important,
{\sf S} should check the user's credential valid period $VP_{U}$ and make sure
that  the ticket valid period $VP_{T}$ is no later than $VP_{U}$. $T_U$ together
with its associated  details is then sent back to the user who can use the
information together with the public key of the seller, $Y_S$, to verify the
validity of the information. Note that our scheme provides {\em ticket
unlinkability} due to the use of user pseudonyms which prevents the seller, {\sf
S}, as well as any verifier, $V$ from linking any two ticket requests by the
same user even if they collude. %
\medskip

\begin{figure*}[!h]
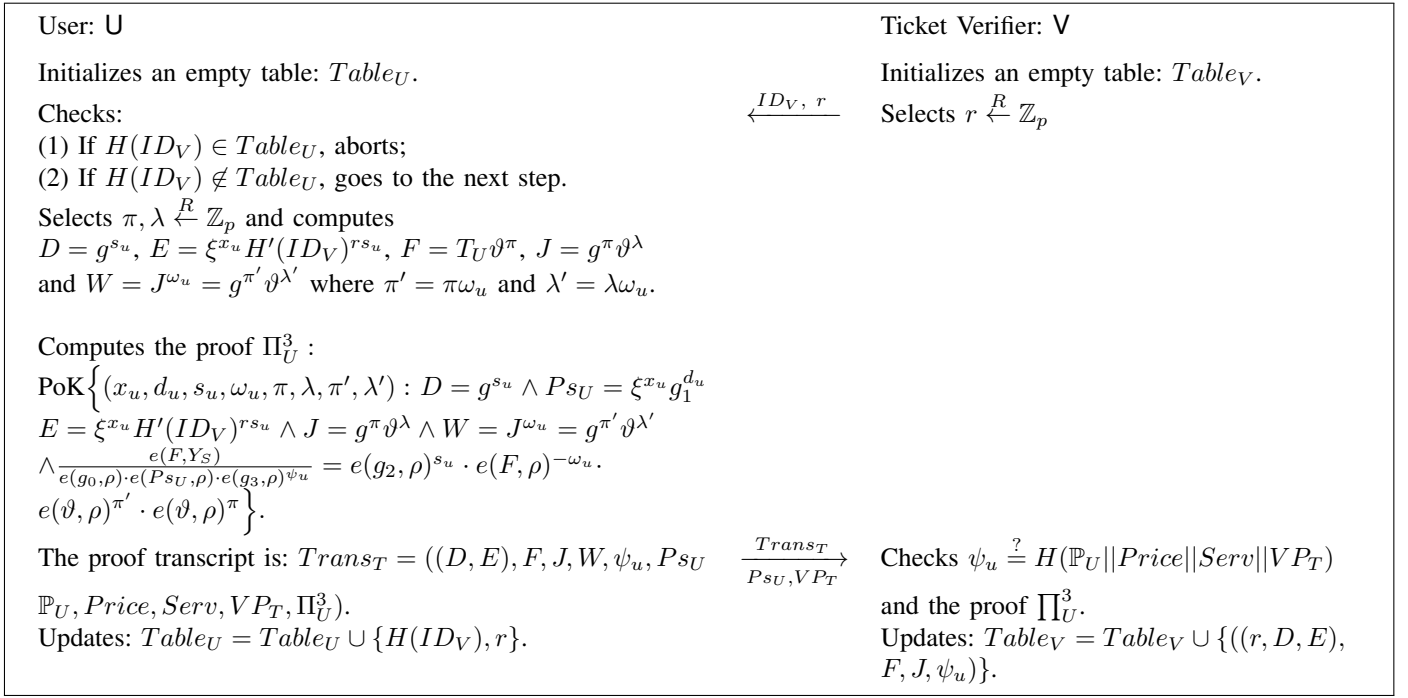

\centering
\fbox{
\begin{minipage}{18cm}

\begin{tabular}{lcl}
User: {\sf U} & & Ticket Verifier: {\sf V}
\medskip\\
Initializes an empty table: $Table_{U}$. & &Initializes an empty table: $Table_{V}$.\\
Checks: & $\xleftarrow{ID_{V},~r~}$ & Selects $r\stackrel{R}{\leftarrow}\mathbb{Z}_{p}$\\
 (1) If $H(ID_{V})\in Table_{U}$, aborts;\\
(2) If $H(ID_{V})\not\in Table_{U}$, goes to the next step.\\
 Selects $\pi,\lambda\stackrel{R}{\leftarrow}\mathbb{Z}_{p}$ and computes\\
$D=g^{s_{u}}$, $E=\xi^{x_{u}}H'(ID_{V})^{rs_{u}}$, $F=T_{U}\vartheta^{\pi}$,  $J=g^{\pi}\vartheta^{\lambda}$ \\
and $W=J^{\omega_{u}}=g^{\pi'}\vartheta^{\lambda'}$ where $\pi'=\pi \omega_{u}$ and $\lambda'=\lambda \omega_{u}$.
\\
&&\\
Computes the proof $\Pi_{U}^{3}:$\\
$ \mbox{PoK}\Big\{
(x_{u},d_{u},s_{u},\omega_{u},\pi,\lambda,\pi',\lambda'): $
$D=g^{s_{u}}\wedge Ps_{U}=\xi^{x_{u}}g_{1}^{d_{u}}$\\$
E=\xi^{x_{u}}H'(ID_{V})^{rs_{u}} \wedge J=g^{\pi}\vartheta^{\lambda}\wedge   W=J^{\omega_{u}}=g^{\pi'}\vartheta^{\lambda'}$\\
$ \wedge 
  \frac{e(F,Y_{S})}{e(g_{0},\rho)\cdot e(Ps_{U},\rho)\cdot e(g_{3},\rho)^{\psi_{u}}}= e(g_{2},\rho)^{s_{u}}\cdot e(F,\rho)^{-\omega_{u}}\cdot $\\
  $ e(\vartheta,\rho)^{\pi'}\cdot e(\vartheta,\rho)^{\pi}
\Big\}.$ \\
The proof transcript is:  $Trans_{T}=((D,E),F,J,W,\psi_{u},Ps_{U}$ & $\xrightarrow[Ps_{U},VP_{T}]{Trans_{T}}$ & Checks $\psi_{u}\stackrel{?}{=}H(\mathbb{P}_{U}||Price||Serv||VP_{T})$ \\
$\mathbb{P}_{U},Price,Serv,VP_{T},\Pi_{U}^{3})$.& &   and the proof 
$\prod_{U}^{3}$.\\

Updates: $Table_{U}=Table_{U}\cup\{H(ID_{V}),r\}$. & &Updates: $Table_{V}=Table_{V}\cup\{((r,D,E),$\\
& &$F,J,\psi_{u})\}$.\\
\end{tabular}
\end{minipage}
}
\caption{Ticket Validation Algorithm}\label{valid}
\end{figure*}

\noindent{\em Ticket Validation.}  Figure~\ref{valid} depicts the necessary
steps to validate a ticket. The user {\sf U} initializes an empty table
$Table_{U}$ to store the identity information of any verifier {\sf V}. The
purpose of this table is to ensure that a verifier can only ask for a ticket
once to prevent an honest user from being de-anonymised by a malicious verifier.
The verifier {\sf V}, on the other hand, initializes an empty table $Table_{V}$
to store the authentication transcript from ${\sf U}$ to determine if a ticket
has already been used (\ie double spend detection). The ticket verification
process is started by the verifier sending a fresh nonce $r$ and its identity,
$ID_V$, to the user. It is assume that there is some out-of-band channel which
allows the user to ``authenticate'' the verifier, \eg it is a guard on the train
or a gate at the entrance of the platform, \etc~  {\sf U} first checks that $V$
does not yet have an entry in $Table_{U}$. If an entry exists, {\sf U} aborts
the process to avoid de-anonymisation. Otherwise, she proceeds to send $V$ a
``ticket transcript'', $Trans_T$, of her ticket $Ticket_U$, which includes a zero
knowledge proof of knowledge, $\Pi_U^3$. The transcript should convince $V$ that
she is a legitimate user who is in the possession of a valid ticket
$Ticket_{U}$. Because $Ticket_U$, includes the user's secret key $x_{u}$ as part
of her pseudonym, $Y$, knowledge of which needs to be demonstrated as part of
$\Pi_U^3$, our scheme ensures {\em ticket untransferability} assuming {\sf U}'s
private key has not been compromised. Moreover, the transcript also incorporates
$V$'s nonce $r$ to prevent simple replay attacks. $U$ completes her part in the
validation process by updating her $Table_{U}$ storing $V$'s identity together
with $r$. If $V$ can successfully verify $U$'s ticket transcript, {\sf V} grants
her access to the service and updates $Table_{V}$ with $Trans_T$; otherwise,
{\sf V} denies the request.

\medskip

\begin{figure*}[!h]
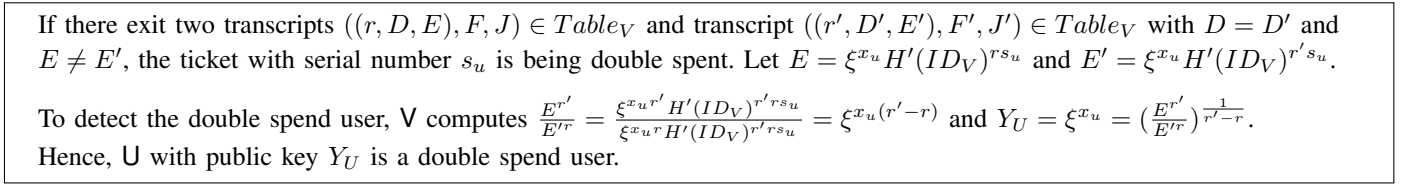

\centering
\fbox{
\begin{minipage}{18cm}
\begin{tabular}{l}

If there exit two transcripts $((r,D,E),F,J)\in Table_{V}$  and transcript $((r',D',E'),F',J')\in Table_{V}$ with $D=D'$ and   \\ $E\neq E'$, the ticket with serial number $s_{u}$ is being double spent. 
\medskip

Let $E=\xi^{x_{u}}H'(ID_{V})^{rs_{u}}$ and $E'=\xi^{x_{u}}H'(ID_{V})^{r's_{u}}$.\\ To detect the double spend user, {\sf V} computes  $\frac{E^{r'}}{E'^{r}}=\frac{\xi^{x_{u}r'}H'(ID_{V})^{r'rs_{u}}}{\xi^{x_{u}r}H'(ID_{V})^{r'rs_{u}}}=\xi^{x_{u}(r'-r)}$ and $Y_{U}=\xi^{x_{u}}=(\frac{E^{r'}}{E'^{r}})^{\frac{1}{r'-r}}$. \\
 Hence, {\sf U} with public key  $Y_{U}$ is a double spend user.\\
\end{tabular}
\end{minipage}
}
\caption{Double Spend Detection}\label{double-spend}
\end{figure*}

\noindent{\em Double Spend Detecting.} Figure~\ref{double-spend} shows the
double spend detection process. To determine whether a ticket is being double
spent, {\sf V} checks $Table_V$ for another ticket transcripts, $Trans_U$, with
the same serial number, $D$. If there is, the ticket is being double spent and
{\sf V} can de-anonymise {\sf U} by extracting her public key, $Y_U$, from the
two transcripts; otherwise, it is a new ticket.

It is worth pointing out that the construction of our scheme has the following 
additional benefit:

\noindent{\em Limited Dynamic Policy Update.} If the  seller {\sf S} needs to
either update some policies in $\mathbb{P}$ or create new ones, he can contact
the central authority {\sf CA} to update or create the relevant public
parameters $params$. As a result, when buying a ticket, a user {\sf U}  proves
to {\sf S} that his attributes satisfy the updated policies by using the updated
$params$ and {\sf S} will generate tickets according using the updated policies.
{\sf U} only needs to obtain new credentials from the {\sf CA} if her current
attributes do not satisfy the updated policies any more. For example, suppose
that Alice is $16$ years old. If the seller {\sf S} requests that the existing
policy range of $[12,18]$ is changed to $[15,20]$ instead, then Alice can still
use her existing credentials. However, if the policy were changed from $[12,18]$
to $[18,25]$ instead, then Alice would need to get the {\sf CA} to update her
credentials.

\medskip

\begin{figure*}[!h]
\hfill
\centering
\fbox{
\begin{minipage}{17.5cm}
{\sf CA} publishes the ticket price polices $\mathbb{P}=\{\mathbb{R}_{1},$
$\cdots,\mathbb{R}_{N_{1}},\mathbb{S}_{1},\cdots,\mathbb{S}_{N_{2}}\}$ 
where $\mathbb{R}_{l}=[c_{l},d_{l}]$ is a range policy (i.e. age, mileage)
 and 
$\mathbb{S}_{i}=\{I_{i_{1}},I_{i_{2}},\cdots,I_{i_{\varsigma}}\}$ 
is a set policy (i.e. location, profession, disbility) and consists of  $\varsigma$ items $I_{i_{j}}$   for  
 $l=1,2,\cdots,N_{1}$ and $i=1,2,\cdots,N_{2}$. 
\medskip\\

{\sf CA} runs $\mathcal{BG}(1^{\ell})\rightarrow(e,p,\mathbb{G},\mathbb{G}_{\tau})$.  Suppose that the longest interval length in $\{\mathbb{R}_{1},\cdots,\mathbb{R}_{N_{1}}\}$   is $[0,q^{k})$ where $q\in\mathbb{Z}_{p}$   and
 $p>2q^{k}+1$.  Let $g,g_{0},g_{1},g_{2},g_{3},\hat{g}_{1},\hat{g}_{2},\cdots,\hat{g}_{N_{1}},h,\mathfrak{g},\eta,\xi,\rho,\vartheta,\eta_{1},\eta_{2},\cdots,\eta_{N_{2}}$ be generators of $\mathbb{G}$, $H:\{0,1\}^{*}\rightarrow\mathbb{Z}_{p}$ and $H':\{0,1\}^{*}\rightarrow\mathbb{G}$  be two cryptographic hash functions.
\medskip\\

  CA selects
  $x,y,\mu_{1},\mu_{2},\cdots,\mu_{N_{2}}\stackrel{R}{\leftarrow}\mathbb{Z}_{p}$
  and computes
  $\tilde{g}=g^{x},\tilde{h}=h^{y},h_{0}=h^{\frac{1}{y}},{h}_{1}=h^{\frac{1}{y+1}},{h}_{2}=h^{\frac{1}{y+2}},\cdots,$
   ${h}_{q-1}=h^{\frac{1}{y+q-1}},$ 
  $\tilde{h}_{0}=h^{q^{0}},\tilde{h}_{1}=h^{q},\cdots,\tilde{h}_{k-1}=h^{q^{k-1}},$
   $\tilde{\eta}_{1}=\eta_{1}^{\mu_{1}},\tilde{\eta}_{2}=\eta_{2}^{\mu_{2}},\cdots,\tilde{\eta_{N_{2}}}=\eta_{N_{2}}^{\mu_{N_{2}}}$
   and $ \left({\eta}_{i_{1}}=\eta^{\frac{1}{\mu_{i}+H(I_{i_{1}})}}, 
  {\eta}_{i_{2}}=\eta^{\frac{1}{\mu_{i}+H(I_{i_{2}})}},\cdots,{\eta}_{i_{\varsigma}}=
   \eta^{\frac{1}{\mu_{i}+H(I_{i_{\varsigma}})}}\right)_{i=1}^{N_{2}}.$ 
  \medskip\\
 
The secret key of CA is $MSK=(x,y,\mu_{1},\mu_{2},\cdots,\mu_{N_{2}})$ and the public parameters are 
$params =(e,p,\mathbb{G},\mathbb{G}_{\tau},g,g_{0},$\\ $g_{1},g_{2},\hat{g}_{1},\hat{g}_{2},\cdots,\hat{g}_{N_{1}},h,\mathfrak{g},\eta,\xi,\rho,\tilde{g},\tilde{h},{h}_{0},{h}_{1},\cdots,{h}_{q-1},\tilde{h}_{0},\tilde{h}_{1},\cdots,\tilde{h}_{k-1}, \eta_{1},\eta_{2},\cdots,$$\eta_{N_{2}},
({\eta}_{i_{1}},{\eta}_{i_{2}},\cdots,$ ${\eta}_{i_{\varsigma}})_{i=1}^{N_{2}}).$\\
\end{minipage}
}
\caption{Setup Algorithm}\label{setup}
\end{figure*}

\begin{figure*}[!h]
\centering
\fbox{
\begin{minipage}{17.5cm}
\medskip

\begin{tabular}{lcl}
Ticket Seller {\sf S} && Central Authority {\sf CA}\\
Selects $x_{s}\stackrel{R}{\leftarrow}\mathbb{Z}_{p}$ and computes $Y_{S}={\rho}^{x_{s}}$. & &\\
Computes the proof $\Pi_{S}^{1}:\mbox{PoK}\{x_{s}:Y_{S}=\rho^{x_{s}}\}$. & $\xrightarrow{ID_{S},Y_{S},\Pi_{S}^{1}}$ & Selects $c_{s},r_{s}\stackrel{R}{\leftarrow}\mathbb{Z}_{p}$  and computes \\
Verifies $e(\sigma_{S},\tilde{g}g^{c_{s}})\stackrel{?}{=}e(g_{0},g)\cdot e(g,g_{1})^{H(VP_{S})}$& $\xleftarrow[~\sigma_{S},~VP_{S}]{c_{s},~r_{s},}$ & $\sigma_{S}=(g_{0}g_{1}^{H(VP_{S})}Y_{S}\mathfrak{g}^{r_{s}})^{\frac{1}{x+c_{s}}}$, where $VP_{S}$ is a valid \\
\hspace{1cm} $\cdot e(Y_{S},g)\cdot e(g,\mathfrak{g})^{r_{s}}.$  & &  period.\\

Keeps the credential $Cred_{S}=(c_{s},r_{s},\sigma_{S})$. & &\\

\vspace{0.2cm}\\

 User {\sf U} & &Central Authority {\sf CA}\\
 Selects $x_{u}\stackrel{R}{\leftarrow}\mathbb{Z}_{p}$ and computes $Y_{U}={\xi}^{x_{u}}$.&&\\
 Selects $r\stackrel{R}{\leftarrow}\mathbb{Z}_{p}$ and compute $R=\mathfrak{g}^{r}$. &&\\

 Computes the proof $\Pi_{U}^{1}:$ &  & \\
  $\mbox{PoK}\left\{(x_{u},r):Y_{U}=\xi^{x_{u}}~\wedge~R=\mathfrak{g}^{r}\right\}$.   &    $\xrightarrow[A_{U},\Pi_{U}^{1}]{ID_{U},Y_{U},R,}$  
 & Selects $c_{u},r'\stackrel{R}{\leftarrow}\mathbb{Z}_{p}$ and computes $\sigma_{U}=$\\
 Computes $r_{u}=r+r'$.   &    $\xleftarrow[\sigma_{U},~VP_{U}]{c_{u},~r',}$& $\left(g_{0}g_{1}^{H(VP_{U})}Y_{U}R\mathfrak{g}^{r'}\prod_{l=1}^{N_{1}}\hat{g}_{l}^{a_{l}}\prod_{i=1}^{N_{2}} \eta_{i}^{H(I_{i_{j}})}\right)^{\frac{1}{x+c_{u}}}$\\

 Verifies 
 $e(\sigma_{U},\tilde{g}g^{c_{u}})=e(g_{0},g)\cdot e(g,g_{1})^{H(VP_{U})}\cdot $ & &  where $VP_{U}$ is a valid period, $a_{l}\in A_{U}\models \mathbb{R}_{l}$ and \\
\hspace{1cm} $e(Y_{U},g)\cdot e(\mathfrak{g},g)^{r_{u}}\cdot  \prod_{l=1}^{N_{1}}e(\hat{g}_{l},g)^{a_{l}} \cdot$ & & $A_{U}\models I_{i_{j}}$.\\
\hspace{1cm}$  \prod_{i=1}^{N_{2}} e(\eta_{i},g)^{H(I_{i_{j}})}$.  & & \\

 Keeps the credential $Cred_{U}=(c_{u},r_{u},\sigma_{U})$.   & & Stores $(ID_{U},A_{U},Y_{U},\sigma_{U})$.\\

  \end{tabular}
\end{minipage}
}
  \caption{Registration Algorithm}\label{regist}
\end{figure*}

\section{Security Analysis}\label{secu}

The correctness of our scheme is shown in Appendix \ref{app_correct}. To
demonstrate its security, we need to prove indistinguishability between the
behaviours of the real-world adversary $\mathcal{A}$ and the behaviours of the
ideal-world adversary $\mathcal{A}'$. Given a real-world adversary
$\mathcal{A}$, there exist an ideal-world adversary $\mathcal{A}'$ such that no
environment $\mathcal{E}$ can distinguish whether it is interacting with
$\mathcal{A}$ or $\mathcal{A}'$.  The proof is based on sublemmas where
different corrupted parties are considered. The following cases are not
considered: (1) the {\sf CA} is the only honest party; (2) the {\sf CA} is the
only dishonest party; (3) all parties are dishonest; and (4) all parties are
honest. The first three do not make a sensible system while the last one is
trivially secure. Since the {\sf CA} needs to know {\sf U}'s attributes to issue
her with her credentials, we assume that {\sf CA} is honest and fully trusted by
the other entities in the system.

In order to prove the indistinguishability between ${\bf Real}_{\mathcal{E},\mathcal{A}}(\ell)$ and ${\bf Ideal}_{\mathcal{E},\mathcal{A}'}$, a sequence of games {\bf Game}$_{0}$, {\bf Game}$_{1}$, $\cdots$, {\bf Game}$_{n}$ are defined. For each {\bf Game}$_{i}$,  we construct a simulator $Sim_{i}$ that runs $\mathcal{A}$ as a subroutine and provides $\mathcal{E}$'s view, for $i=0,1,\cdots, n$. ${\bf Hybrid}_{\mathcal{E},Sim_{i}}(\ell)$ denotes the probability that $\mathcal{E}$ outputs $1$ running in the world provided by $Sim_{i}$. $Sim_{0}$ runs $\mathcal{A}$ and other honest parties in the real-world experiment, hence ${\bf Hybrid}_{\mathcal{E},Sim_{0}}={\bf Real}_{\mathcal{E},\mathcal{A}}$. $Sim_{n}$ runs $\mathcal{A}'$ in ideal-world experiment, hence ${\bf Hybrid}_{\mathcal{E},Sim_{n}}(\ell)={\bf Ideal}_{\mathcal{E},\mathcal{A}'}(\ell)$. Therefore,  
\begin{equation*}
\begin{split}
&\left|{\bf Real}_{\mathcal{E},\mathcal{A}}(\ell)- {\bf Ideal}_{\mathcal{E},\mathcal{A}'}\right| \leq 
 \left|{\bf Real}_{\mathcal{E},\mathcal{A}}(\ell)- {\bf Hybrid}_{\mathcal{E},Sim_{1}}\right|\\
&+ \left|{\bf Hybrid}_{\mathcal{E},Sim_{1}}-h{\bf Hybrid}_{\mathcal{E},Sim_{2}}\right|+\cdots
+\\
& \left|{\bf Hybrid}_{\mathcal{E},Sim_{n-1}}-{\bf Hybrid}_{\mathcal{E},Sim_{n}}\right|.
\end{split}
\end{equation*}
\begin{theorem}\label{theorem}
Our privacy-preserving electronic ticket scheme   with attribute-based
credentials described in Fig. \ref{setup}, Fig. \ref{regist}, Fig. \ref{issue},
Fig. \ref{valid} and Fig. \ref{double-spend} is secure if the $q$-strong
Diffie-Hellamn assumption ($q$-SDH) holds on the bilinear group
$(e,p,\mathbb{G},\mathbb{G}_{\tau})$.
\end{theorem}

Theorem \ref{theorem} is proven by using the following two lemmas.

\begin{lemma}\label{l1} {\sf (User Privacy)} 
For all environments $\mathcal{E}$ and all real-world adversaries $\mathcal{A}$
who statically control the ticket seller  {\sf S} and verifier {\sf V}, there
exists an ideal-word adversary $\mathcal{A}'$ such that $ \left|{\bf
Real}_{\mathcal{E},\mathcal{A}}(\ell)-{\bf
Ideal}_{\mathcal{E},\mathcal{A}'}(\ell)\right|\leq 2^{\frac{1}{\ell}}. $
\end{lemma}

The proof of {\em Lemma \ref{l1}} is given in Appendix \ref{p1}. Since 
anonymity, ticket unlinkability and ticket untransferability are part of user
privacy, they are therefore proved by  {\em Lemma \ref{l1}}.

\begin{lemma} \label{l2}{\sf (Seller Security)} 
For all environments $\mathcal{E}$ and all real-world adversaries $\mathcal{A}$
who statically controls the verifier {\sf V} and one or more users, there exists
an ideal-word adversary $\mathcal{A}'$ such that

\begin{equation*}
\begin{array}{ll}
\left|{\bf Real}_{\mathcal{E},\mathcal{A}}(\ell)-{\bf
Ideal}_{\mathcal{E},\mathcal{A}'}(\ell)\right|
\leq\frac{q_{T}}{2^{\ell}}+\frac{q_{v}}{2^{\ell}}+\frac{1}{q_{I}}Adv_{\mathcal{A}}^{q_{I}\mbox{-SDH}}(\ell)
 \\ 
+Adv_{\mathcal{A}}^{(q+1)\mbox{-SDH}}(\ell)+Adv_{\mathcal{A}}^{(\varsigma+1) 
\mbox{-SDH}}(\ell)+\frac{1}{q_{T}}Adv_{\mathcal{A}}^{q_{T}\mbox{-SDH}}(\ell)\\
 +\frac{1}{q_{V}}Adv_{\mathcal{A}}^{q_{V}\mbox{-SDH}}(\ell),
\end{array}
\end{equation*}
where $q_{T}$, $q_{I}$, $q_{V}$ are the number of ticket issue queries, credential queries  and ticket validation queries made by $\mathcal{A}$, respectively.
\end{lemma}

The proof of {\em Lemma \ref{l2}} is given in Appendix \ref{p2}. Since
unforgeability, double spending detection and de-anonymization are included in
the seller security, they are therefore proved by {\em Lemma \ref{l2}}.
Therefore, {\em Theorem 4} is proven because both {\em Lemma \ref{l1}} and {\em
Lemma \ref{l2}} hold.

\section{Benchmarking results} 
\label{sec:benchmarks}

In this section we evaluate the performance of our scheme. The source code of
the scheme's implementation is available at \cite{githubrepo} and its
performance has been measured on a Dell Inspiron Latitude E5270 laptop with an
Intel Core i7-6600U CPU, 1TB SSD and 16GB of RAM running Fedora 27. %
The implementation makes use of bilinear maps defined over elliptic curves as
well as other cryptographic primitives. We used the JPBC library\cite{jpbc} for
the bilinear maps and bouncycastle\cite{bouncycastle} for the other
cryptographic required by our scheme. Note that the Java based implementation of
the JPBC API\cite{jpbc} was used throughout.

Recall from Section~\ref{prelim} that our scheme requires a Type I symmetric
bilinear map, $e:\mathbb{G} \times \mathbb{G} \rightarrow \mathbb{G}_{\tau}$.
The JPBC library \cite{jpbc} provides three different instances of a symmetric
pairing with their Type A, A1 or E pairings. The Type A and A1 pairings are
based on the  elliptic curve $E:y^{2}=x^{3}+x$ over the finite field $F_p$. In
both cases, the group $\mathbb{G}$ in  is the group of points on the elliptic
curve, $E(F_p)$. The Type E pairing, on the other hand, is based on the Complex
Multiplication (CM) method of constructing elliptic curves starting with the
Diophantine equation $DV^2=4p-t^2$. The details of each construction can
be found in \cite{lynn2007implementation}.

In our implementation, we use the default parameters during the instantiation of
the different pairings, \eg Type A is constructed using $rBits=160$, $qBits=512$,
Type A1 uses $2$ primes of size $qBits=512$ and Type E is instantiated with
$rBits=160$ and $qBits=1024$.


Note that according to Table 1 in \cite{freeman2010taxonomy}, JPBC's default
Type A pairing provides approximately the equivalent of $80$-bit symmetric or
$1024$ RSA-style security. This is sufficient for providing a baseline for
taking time measurements.




For the hash functions $H:\{0,1\}^{*}\rightarrow\mathbb{Z}_{p}$ and
$H':\{0,1\}^{*}\rightarrow\mathbb{G}$ required by our scheme (see
Fig~\ref{setup}), we used $SHA256$ for $H$ and rely on the implementation of
``newElementFromHash()" method in the JPBC library for $H'$.


\subsection{Timings} 

Table~\ref{tab:Benchmarks} shows the results of the computational time spent in
the various phases of our proposed scheme which required more complex
computations (\ie some form of verification using bilinear maps or generation of
zero knowledge proofs). The timings shown have been calculated as the average 
over $20$ iterations.  

The maximum range interval in this instance was $7$ which is covered by the
interval $[0,2^3)$ and thus $k=3$ in the set-up algorithm described in
Fig~\ref{setup}. The maximum set size used was $10$. It is clear from the
computations involved in the generation of $\Pi_U^2$ (see
Appendix~\ref{app:PI_U_2}) that the computational cost of a range proof
increases with $k$ while the number of computations for a set membership proof
is independent on the size of the set. As such the numbers presented below
provide a reasonable lower bound of the computational costs for range proofs
assuming that any useful ranges will have at least an interval length of $4$ or
more.

\begin{table*}\caption{Benchmark results {(in ms)}} \label{tab:Benchmarks}
\centering{
\begin{tabular} {|l|c|c|c|c|}  
\hline
Protocol phase &Entity&\multicolumn{3}{|c|}{(\#range policies,\#set 
policies)$=(2,4)$}\\
&&Type A & Type A1 & Type E\\
\hline
\multicolumn{5}{|c|}{System Initialisation -  Central Authority 
($\mathcal{CA}$)}\\
\hline
initialise the system &CA &{$626.05.1$} &$9155.95$ &$2895.25$ \\
\hline

\multicolumn{5}{|c|}{Issuing phase}\\
\hline
generate PoK $\Pi_S^2$ &Seller &{$184.25$} &$2881.8$ & $469.1$\\
\hline
verify $\Pi_S^2$ &User &{$107.9$}&$1424.95$ & $286.2$\\ 
\hline
generate ticket request, $\Pi_U^2$ &User &$1008.7$ &$17195.95$ &$2847.35$ \\
\hline
verify $\Pi_U^2$ & Seller &$787.3$ &$11288.0$ &$2166.25$ \\ 
\hline
generate ticket & Seller &{$47.85$} &$583.0$ &$120.3$ \\
\hline
verify ticket  &User &{$52.5$}&$856.75$ &$158.35$ \\
\hline
\multicolumn{5}{|c|}{Ticket Verification - Verifier ($\mathcal{V}$)}\\
\hline
generate ticket transcript $Trans_T$ &User &{$241.4$} &$3538.7$ &$707.4$ \\ 
\hline
verify transcript & Verifier &{$214.05$}&$2539.8$ & $649.8$ \\
\hline
\multicolumn{5}{|c|}{Total system run time}\\
\hline
All phases&All &$3659.1$ &$54382.95$ & $11383.1$\\
\hline 
\end{tabular}
}\medskip
\end{table*}

\begin{table*}\caption{Type A: benchmark results for different ranges and set sizes {(in 
ms)}}\label{tab:Benchmarks2}
\centering{
\begin{tabular}{|l|c|c|c|c|c|}
\hline
Ticket issuing phase &$k=5$ &$k=10$ &$k=20$ &$s=10$ &$s=100$\\
                     &$[0,31]$ &$[0,1023]$ &$[0,1048755]$ &$\{x|1\leq x\leq 
                     10\}$ &$\{x|1\leq x\leq 100\}$\\
\hline
range/set proof creation &$\approx 512$ &$\approx 961$&$\approx 1998$&$\approx 
35$&$\approx 36$ 
\\
\hline
range/set proof verification &$\approx 367$ &$\approx 599$ &$\approx 
1116$ &$\approx 22$&$\approx 23$\\ 
\hline
\end{tabular}
}\medskip
\end{table*}


Table~\ref{tab:Benchmarks} shows the timings for our current implementation of
our scheme with $2$ small range policies and $4$ set policies using the default
instantiation of the three different symmetrical pairings available in JPBC. The
fastest performance is achieved by the JPBC Type A curved based Type I pairing,
where ticket issuing and verification take $\approx 2.2$s and $\approx 450$ms
respectively. %

Table~\ref{tab:Benchmarks2} illustrates the impact of different range and set
sizes on the computational effort during the ticket issuing phase using the 
JBPC Type A curve. It is clear that set membership proofs can be computed much 
faster than range proofs and their computational cost is independent of the set size whereas for range proofs the computational effort increases linearly with $k$.

However, range proofs provide an additional benefit which is best illustrated
with an example: a young person's age could be either codified in a range policy
($age\in[15,25]$) or with a set policy (``young person''). Our scheme provides
the policy maker with the flexibility to decide which kind of policies should be
used. While a set policy is computationally more efficient than a range policy,
range policies can potentially accommodate future policy changes. In particular,
suppose that Alice is $23$ years old and the current young person range policy
is given by $age\in[16,22]$ which means Alice cannot obtain a discount. However,
if it is later changed to $age\in[16,25]$, Alice can still use her existing age
attribute of $23$ to obtain a young person discount as she can now prove her age
falls within the updated range. However, if the set policy approach had been
used, Alice would need to return to the $CA$ to update her credentials as she
would not have been eligible for her signed ``young person'' attribute,
previously.

Consequently, for any real system, it is important to look at the trade-off
between the flexibility that range policies allow in terms of dynamic updates
and their more expensive computational cost.
 \medskip
 
Note that our implementation has not yet been optimised and thus it should be
possible to improve its performance considerably by pre-computing static values
off-line where possible and switching from the current Java-based version to
using a Java-wrapper to the C-based implementation of the PBC 
libraries~(\cite{pbc:2006}), instead. This might also go some way to ameliorate
the computational burden of range policies.

\section{Future Work and Conclusions}%
\label{conc}%

To protect user privacy in e-ticket schemes, various schemes have been proposed
but which did not address attribute-based ticketing. This paper presented a
scheme  which implements attribute-based ticketing while protecting user
privacy. Our proposed scheme makes the following contributions:
(1) users can buy different tickets from ticket sellers without releasing their
exact attributes; (2) two tickets of the same user cannot be linked; (3) a
ticket cannot be transferred to another user; (4) a ticket cannot be double
spent; (5) the security of the proposed scheme is formally proven and reduced to
well-known ($q$-strong Diffie-Hellman) complexity assumption; (6) the scheme has
been implemented and its performance empirical evaluated. 
\medskip

Our future work will be looking at the impact on the security model and proof
when dynamic policy updates are allowed as well as changes to scheme's
implementation to improve its performance, \eg by pre-computing static values
where possible and using the C-based PBC library\cite{pbc:2006}, instead.
\medskip


\appendix{}\label{app}
\subsection{Correctness}\label{app_correct}
We claim that our  scheme described in Fig. \ref{setup}, Fig.
\ref{regist}, Fig. \ref{issue}, Fig. \ref{valid} and Fig. \ref{double-spend} is
correct. Using the properties of bilinear maps, it is trivial to validate that
the equations in Fig. \ref{regist}, which are used to verify the credentials
sent to the seller and user, hold. Similarly, it is straightforward to verify
the equations in Fig. \ref{double-spend}, which are used by ticket verifiers to
detect double spending and de-anonymize users who have double spent tickets.

We will now demonstrate that the correctness of the equations in Fig.
\ref{issue} and Fig. \ref{valid} also hold true. The former are used by  a user
and a ticket seller to prove that they hold valid credentials, while the latter
equations  are used by a user to prove that she holds a valid ticket issued a
ticket seller.

To verify the seller proof $\Pi_S^2$, we use the following equality:
\begin{equation}\label{e1}
\begin{split}
&e(Q,\tilde{g})=e(\sigma_{S}\vartheta^{z},g^{x})\\
&=e((g_{0}g_{1}^{H(VP_{S})}Y_{S}\mathfrak{g}^{r_{s}})^{\frac{x+c_{S}-c_{S}}{x+c_{S}}},g)\cdot e(\vartheta,\tilde{g})^{z}\\
&=e((g_{0}g_{1}^{H(VP_{S})}Y_{S}\mathfrak{g}^{r_{s}}),g)\cdot e(\sigma_{S}\vartheta^{z},g)^{-c_{S}} \cdot \\
& ~~~~e(\vartheta,g)^{-c_{S}z}\cdot  e(\vartheta,\tilde{g})^{z}\\
&=e(g_{0},g)e(g_{1},g)^{H(VP_{S})}\cdot e(\rho,g)^{x_{S}}\cdot e(\mathfrak{g},g)^{r_{s}}\cdot  \\
 &~~~~ e(Q,g)^{-c_{S}} \cdot e(\vartheta,g)^{-c_{S}z}\cdot e(\vartheta,\tilde{g})^{z}.
\end{split}
\end{equation}

Hence, we have

\begin{equation}\label{e2}
\begin{split}
&\frac{e(Q,\tilde{g})}{e(g_{0},g)\cdot e(g_{1},g)^{H(VP_{S})}}=e(\rho,g)^{x_{S}}\cdot e(\mathfrak{g},g)^{r_{s}}\cdot \\
& e(Q,g)^{-c_{S}} \cdot e(\vartheta,g)^{-c_{S}z}\cdot  e(\vartheta,\tilde{g})^{z}.
\end{split}
\end{equation}
Eqs. (\ref{e1}) and (\ref{e2}) are used to show the correctness of the proof of the ticket issuer's credential. 
\medskip

Let $\Delta=g_{0}\cdot{e(g_{1},g)^{H(VP)}}Y_{U}\mathfrak{g}^{r_{u}}\prod_{l=1}^{N_{1}}\hat{g}_{l}^{a_{l}}\prod_{i=1}^{N_{2}}\eta_{i}^{H(I_{i_{j}})}$.
 \begin{equation}\label{e3}
\begin{split}
&e(C,\tilde{g})=e\left(\Delta^{\frac{1}{x+c_{u}}}\vartheta^{\alpha},g^{x}\right)
=e\left(\Delta^{\frac{x+c_{u}-c_{u}}{x+c_{u}}},g\right)\cdot e(\vartheta,\tilde{g})^{\alpha}\\
&=e\left(\Delta,g\right)\cdot e\left(\Delta^{\frac{-c_{u}}{x+c_{u}}}\vartheta^{-c_{u}\alpha},g\right)\cdot e(\vartheta,g)^{c_{u}\alpha}\cdot e(\vartheta,\tilde{g})^{\alpha}\\
& =e\left(\Delta,g\right)\cdot e\left(C,g\right)^{-c_{u}}\cdot e(\vartheta,g)^{c_{u}\alpha}\cdot e(\vartheta,\tilde{g})^{\alpha}\\
&= e(g_{0},g)\cdot e(g_{1},g)^{H(VP)}\cdot e(\xi,g)^{x_{u}}\cdot e(g,\mathfrak{g})^{r_{u}}\cdot\\
&\hspace{0.3cm} \prod_{l=1}^{N_{1}} e(\hat{g}_{l},g)^{a_{l}}\cdot  \prod_{i=1}^{N_{2}}e(\eta_{i},g)^{H(I_{i_{j}})}\cdot e(C,g)^{-c_{u}}\cdot \\
 &  \hspace{0.3cm}  e(\vartheta,g)^{c_{u}\alpha}\cdot  e(\vartheta,\tilde{g})^{\alpha}.
\end{split}
\end{equation}

Hence,

\begin{equation}\label{e4}
\begin{split}
 &   \frac{e(C,\tilde{g})}{e(g_{0},g)\cdot e{g_{1},g}^{H(VP_{U})}}= e(\xi,g)^{x_{u}}\cdot e(g,\mathfrak{g})^{r_{u}}\cdot \prod_{l=1}^{N_{1}}
e(\hat{g}_{l},g)^{a_{l}}\cdot \\ 
 &  \prod_{i=1}^{N_{2}}e(\eta_{i},g)^{H(I_{i_{j}})}\cdot e(C,g)^{-c_{u}}\cdot e(\vartheta,g)^{c_{u}\alpha}\cdot e(\vartheta,\tilde{g})^{\alpha}.
  \end{split}
\end{equation}

   \begin{equation}\label{e5}
\begin{split}
&e(A_{w_{i}},\tilde{h}) =e(h^{\frac{t_{i}}{y+w_{i}}},h^{y})=e(h^{\frac{t_{i}y+t_{i}w_{i}-t_{i}w_{i}}{y+w_{i}}},h)\\
& =e(h,h)^{t_{i}}\cdot e(h^{\frac{t_{i}}{y+w_{i}}},h)^{-w_{i}}=e(h,h)^{t_{i}}\cdot e(A_{i},h)^{-w_{i}}.\\
\end{split}
\end{equation}

\begin{equation}\label{e6}
\begin{split}
&e(A'_{w_{i}},\tilde{h}) =e(h^{\frac{t'_{i}}{y+w'_{i}}},h^{y})=e(h^{\frac{t'_{i}y+t'_{i}w_{i}-t'_{i}w'_{i}}{y+w'_{i}}},h)\\
&=e(h,h)^{t'_{i}}\cdot e(h^{\frac{t'_{i}}{y+w_{i}}},h)^{-w'_{i}} =e(h,h)^{t'_{i}}\cdot e(A_{i},h)^{-w'_{i}}.
\end{split}
\end{equation}

\begin{equation}\label{e7}
\begin{split}
e(B_{i_{j}},\tilde{\eta}_{i})&=e(\eta_{i_{j}}^{e_{i}},\eta_{i}^{\mu_{i}})=e(\eta^{\frac{e_{i}}{\mu_{i}+H(I_{i_{j}})}},\eta^{\mu_{i}})\\
&=e(\eta^{\frac{e_{i}\mu_{i}}{\mu_{i}+H(I_{i_{j}})}},\eta)=e(\eta^{\frac{e_{i}(\mu_{i}+H(I_{i_{j}})-e_{i}H(I_{i_{j}})}{\mu_{i}+H(I_{i_{j}})}},\eta)\\
&=e(\eta,\eta)^{e_{i}}\cdot e(B_{i_{j}},\eta)^{H(I_{i_{j}})}.
\end{split}
\end{equation}
Eqs. (\ref{e3}), (\ref{e4}), (\ref{e5}), (\ref{e6}) and (\ref{e7}) are used to 
show the correctness of the proof of a user's credential and his attributes 
satisfying the ticket policies.
\medskip

\begin{equation}\label{e8}
\begin{split}
&e(F,Y_{S})
=e((g_{0}Yg_{1}^{d_{2}}g_{2}^{s_{u}}g_{3}^{\psi_{u}})^{\frac{1}{x_{s}+\omega_{u}}}\vartheta^{\pi},\rho^{x_{s}})\\
&=e((g_{0}\xi^{x_{u}}g_{1}^{d}g_{1}^{d'}g_{2}^{s_{u}}g_{3}^{\psi_{u}})^{\frac{x_{s}+\omega_{u}-\omega_{u}}{x_{s}+\omega_{u}}},\rho)\cdot e(\vartheta,Y_{S})^{\pi}\\
&=e(g_{0}Ps_{U}g_{2}^{s_{u}}g_{3}^{\psi_{u}},\rho)\cdot e((g_{0}Ps_{U}g_{2}^{s}g_{3}^{\psi_{u}})^{\frac{-\omega_{u}}{x_{s}+\omega_{u}}}, \rho)\cdot e(\vartheta,Y_{S})^{\pi}\\
&=e(g_{0},\rho)\cdot e(Ps_{U},\rho)\cdot e(g_{2},\rho)^{s_{u}}\cdot e(g_{3},\rho)^{\psi_{u}}\cdot\\
& e((g_{0}Ps_{U}g_{2}^{s_{u}}g_{3}^{\psi_{u}})^{\frac{-\omega_{u}}{x_{s}+\omega_{u}}}\vartheta^{-\omega_{u}\pi},\rho)\cdot e(\vartheta,\rho)^{\omega_{u}\pi}\cdot e(\vartheta,Y_{S})^{\pi}\\
&=e(g_{0},\rho)\cdot e(Ps_{U},\rho)\cdot e(g_{2},\rho)^{s_{u}}\cdot e(g_{3},\rho)^{\psi_{u}}\\
&e(F,\rho)^{-\omega_{u}}\cdot  e(\vartheta,\rho)^{\omega_{u}\pi}\cdot e(\vartheta,Y_{S})^{\pi}.
\end{split}
\end{equation}

 Hence,
 
\begin{equation}\label{e9}
\begin{split}
&\frac{e(F,Y_{S})}{e(g_{0},\rho)\cdot e(Ps_{U},\rho)\cdot e(g_{3},\rho)^{\psi_{u}}}=  e(g_{2},\rho)^{s_{u}}\cdot  \\
 & e(F,\rho)^{-\omega_{u}}\cdot e(\vartheta,\rho)^{\omega_{u}\pi}\cdot e(\vartheta,Y_{S})^{\pi}.
\end{split}
\end{equation}

Eqs. (\ref{e8}) and (\ref{e9}) are used to show the correctness of a ticket
proof generated by a user.

\subsection{Proof of Lemma \ref{l1}}\label{p1}
\begin{proof}
To simplify this proof, let {\sf U} be a single honest user since $\mathcal{A}$ can simulate other users by himself.

\noindent{\bf Game-1.}  When $\mathcal{E}$ first makes a ticket-issuing query, the simulator ${Sim}_{1}$  runs the extractor of the proof of knowledge 
\begin{equation*}
\mbox{PoK}\left\{\begin{array}{l} (c_{s},r_{s},\sigma_{S},z,v):
 Z=g^{z}\vartheta^{v} \wedge Z^{c_{s}}=g^{z'}\vartheta^{v'}\\
 \wedge
\frac{e(Q,\tilde{g})}{e(g_{0},g)\cdot e(g_{1},g)^{H(VP_{S})}}=  e(\rho,g)^{x_{s}}\cdot e(\mathfrak{g},g)^{r_{s}}\\
\cdot e(Q,g)^{-c_{s}}\cdot e(\vartheta,g)^{c_{s}z}\cdot
 e(\vartheta,\tilde{g})^{z}
\end{array}
\right\}.
\end{equation*}
\noindent to extract from $\mathcal{A}$ the knowledge $(x_{s},c_{s},r_{s},\sigma_{S},z,v)$ such that $Y_{S}=\rho^{x_{s}}$, $\sigma_{S}=(g_{0}Y_{s}\mathfrak{g}^{r_{s}})^{\frac{1}{x+c_{s}}}$, $Q=\sigma_{S}\vartheta^{z}$, $Z=g^{z}\vartheta^{v}$ and $Z^{c_{s}}=g^{zc_{s}}\vartheta^{vc_{s}}$. If the extractor fails, ${\sf Sim}_{1}$ returns $\mathcal{E}$ with $\perp$ to show the failure; otherwise, ${\sf Sim}_{1}$ runs $\mathcal{A}$ to interact with the honest user. The difference between ${\bf Hybrid}_{\mathcal{E},{Sim}_{0}}$ and ${\bf Hybrid}_{\mathcal{E},{Sim}_{1}}$ lies in the knowledge error of the proof of knowledge, hence 
$\left|{\bf Hybrid}_{\mathcal{E},{ Sim}_{0}} - {\bf Hybrid}_{\mathcal{E},{Sim}_{1}}\right|\leq 2^{\frac{1}{\ell}}.$
\medskip

\noindent{ \bf Game-2.} The simulator $Sim_{2}$ works exactly as $Sim_{1}$ except that it lets the honest user {\sf U} to query a ticket for which his attributes $A_{U}$ $(A_{U}\models \mathbb{P})$.  Due to the (perfect) zero-knowledgeness of the proof of knowledge, $Sim_{2}$ generates a simulated proof of knowledge: 
{\small
\begin{equation*}
\mbox{PoK}\left\{\begin{array}{l l} \big(x_{u}, c_{u},r_{u}, d,\alpha,\beta,\alpha',\beta', (a_{l}, (t_{l_{i}},t'_{l_{i}},w_{l_{i}},\\
w'_{l_{i}})_{i=0}^{k-1})_{l=1}^{N_{1}}, (e_{i}, H(I_{i_{j}}))_{i=1}^{N_{2}}\big):Y=\xi^{x_{u}}g_{1}^{d} \\ 
\wedge Z_{l}=g^{\gamma}h^{a_{l}}  
\wedge D=g^{\alpha}\vartheta^{\beta}~\wedge~D^{c_{u}}=g^{\alpha'}\vartheta^{\beta'}\\
 \wedge \frac{e(C,\tilde{g})}{e(g_{0},g)\cdot e(g_{1},g)^{H(VP_{U})}}=e(\xi,g)^{x_{u}}
   \cdot e(\mathfrak{g},g)^{r_{u}}\cdot\\ 
   \prod_{l=1}^{N_{1}}e(\hat{g}_{l},g)^{a_{l}} \cdot \prod_{i=1}^{N_{2}}\
e(\eta_{i},g)^{H(I_{i_{j}})} \cdot 
   \\ e(\sigma,g)^{-c_{u}}\cdot  e(\vartheta,g)^{\alpha'}\cdot
      e(\vartheta,\tilde {g})^{\alpha}  \wedge~ \big(Z_{l}h^{-c_{l}}=\\
g^{\gamma_{1}}\prod_{i=0}^{k-1}\tilde{h}_{i}^{w_{l_{i}}}
 \wedge Z_{l}\hat{g}_{1}^{-(d_{l}-q^{k})}=g^{\gamma_{1}}\prod_{i=0}^{k-1}\tilde{h}_{i}^{w'_{_{l_{i}}}}   \\
 \wedge (e(A_{w_{l_{i}}},\tilde{h})=e(h,h)^{t_{l}}\cdot e(A_{w_{l_{i}}},h)^{-w_{l_{i}}})_{j=0}^{k-1}~\wedge\\
      (e(A'_{w_{l_{i}}},\tilde{h})=e(h,h)^{t_{l}}\cdot e(A'_{w_{l_{i}}},h)^{-w'_{l_{i}}})_{j=0}^{k-1}\big)_{l=1}^{N_{1}}\\
   \wedge~ \left(e(B_{i_{j}},\tilde{\eta}_{i})=e(\eta,\eta)^{e_{i}}\cdot e(B_{i_{j}},\eta)^{H(I_{i_{j}})}\right)_{i=1}^{N_{2}}
\end{array}\right\}.
\end{equation*}}
We have 
$\left|{\bf Hybrid}_{\mathcal{E},Sim_{1}}-{\bf Hybrid}_{\mathcal{E},Sim_{2}}\right|=0.$
\medskip

\noindent{\bf Game-3.} The simulator $Sim_{3}$ runs exactly as $Sim_{2}$, except that it lets the honest user ${\sf U}$ to valid his/her ticket. Due to the (perfect) zero-knowledgeness of the proof of knowledge, $Sim_{3}$ generaes a simulated proof: 
\begin{equation*}
\mbox{PoK}\left\{
\begin{array}{ll}
(x_{u},d_{u},s_{u},\omega_{u},\pi,\lambda,\pi',\lambda'): 
D=g^{s_{u}}\wedge \\
Ps_{U}=\xi^{x_{u}}g_{1}^{d_{u}}\wedge E=\xi^{x_{u}}H'(ID_{V})^{rs_{u}} \\
 J=g^{\pi}\vartheta^{\lambda} \wedge J^{\omega_{u}}=g^{\pi'}\vartheta^{\lambda'}\wedge \\
  \frac{e(F,Y_{S})}{e(g_{0},\rho)\cdot e(Ps_{U},\rho)\cdot e(g_{3},\rho)^{\psi_{u}}}=  e(g_{2},\rho)^{s_{u}}\cdot  \\
e(F,\rho)^{-\omega_{u}}\cdot e(\vartheta,\rho)^{\pi'}\cdot e(\vartheta,\rho)^{\pi}
\end{array}
\right\}. 
\end{equation*}
We have 
$\left|{\bf Hybrid}_{\mathcal{E},Sim_{2}}-{\bf Hybrid}_{\mathcal{E},Sim_{3}}\right|=0.$
\medskip

\noindent{\bf Game-4.} According to the real-world advesary $\mathcal{A}$, we
construct an ideal-world adversary $\mathcal{A}'$ that plays the simultaneous
roles of the seller ${\sf S}'$ and the verifier ${\sf V}'$, and incorporate all
steps from {\sf Game$_{3}$}. $\mathcal{A}'$ runs $\mathcal{A}$ to obtain system
parameters $params$. When receiving a  user  ${\sf U}'$ registration query
$(registration, ID_{U'}, A_{U'})$ from the trusted third party {\sf TP},
$\mathcal{A}'$ executes the side of the user {\sf U} with $\mathcal{A}$. If the
credential is valid, $\mathcal{A}'$ sends $\tilde{v}=1$ to $TP$ and adds $({\sf
U}',ID_{U}, A_{U}')$ to the user credential list (UCL); otherwise,
$\tilde{\nu}=0$ is returned. For the first time that it receives a ticket issue
query $(ticket\_issuing,Ps_{U'},\mathbb{P}_{U'},VP,Serv)$  from {\sf TP}, it
runs $\mathcal{A}$ to obtain the elements $(x_{s},c_{s},r_{s},\sigma_{S},z,v)$.
$\mathcal{A}'$ simulates a honest user {\sf U}'s query on
$(ticket\_issuing,Ps_{U'},\mathbb{P}_{U'},VP_{T},Serv)$. If the ticket is valid,
$\mathcal{A}'$ sends $\hat{\nu}=1$ to {\sf TP} adds $(Ps_{U'},\mathbb{P}_{U'},
VP_{T},Price,Serv)$ to the user ticket list; otherwise, $\hat{\nu}=0$ is
returned to show failure. When receiving a ticket validating query
$(ticket\_validating,T_{U},\mathbb{P}_{U},Price,VP_{T})$  from {\sf TP},
$\mathcal{A}'$ run $\mathcal{A}$ to obtain the transcript $Trans$ of the proof
of knowledge of $(x_{u},d_{u},s,\omega_{u},pi,\lambda,\pi',\lambda')$. If 
$Trans$ is valid, $\mathcal{A}'$ sends a bit $\bar{\nu}=1$ to {\sf TP}  and adds
$(T_{U},Ps_{U'},\mathbb{P}_{U'}, VP_{T},Price,Serv,Trans)$ to the ticket
validation list; otherwise, $\bar{\nu}=0$ is returned. 
$\mathcal{A}'$ provides $\mathcal{E}$ exactly the same environment as $Sim_{3}$,
hence
$\left|{\bf Hybrid}_{\mathcal{E},Sim_{4}}-{\bf Hybrid}_{\mathcal{E},Sim_{3}}\right|=0.$

Therefore,  
$\left|{\bf Hybrid}_{\mathcal{E},Sim_{0}}-{\bf Hybrid}_{\mathcal{E},Sim_{4}}\right|\leq   
2^{\frac{1}{\ell}}.$
\end{proof}

\subsection{Proof of Lemma \ref{l2}}\label{p2}
\begin{proof} Our PPETS-FGP prevents users from pooling their credentials, hence we should consider multiple users. Some of them can be corrupted by $\mathcal{A}$.
\medskip

\noindent{\bf Game-1.} For each ticket issuing query from a corrupted user dictated by $\mathcal{E}$, the simulator $Sim_{1}$ runs the extractor of the proof of knowledge 

{\small
\begin{equation*}
\mbox{PoK}\left\{\begin{array}{l l} \big(x_{u}, c_{u},r_{u}, d,\alpha,\beta,\alpha',\beta', (a_{l}, (t_{l_{i}},t'_{l_{i}},w_{l_{i}},\\
w'_{l_{i}})_{i=0}^{k-1})_{l=1}^{N_{1}}, (e_{i}, H(I_{i_{j}}))_{i=1}^{N_{2}}\big):Y=\xi^{x_{u}}g_{1}^{d} \\ 
\wedge Z_{l}=g^{\gamma}h^{a_{l}}  
\wedge D=g^{\alpha}\vartheta^{\beta}~\wedge~D^{c_{u}}=g^{\alpha'}\vartheta^{\beta'}\\
 \wedge \frac{e(C,\tilde{g})}{e(g_{0},g)\cdot e(g_{1},g)^{H(VP_{U})}}=e(\xi,g)^{x_{u}}
   \cdot e(\mathfrak{g},g)^{r_{u}}\cdot\\ 
   \prod_{l=1}^{N_{1}}e(\hat{g}_{l},g)^{a_{l}} \cdot \prod_{i=1}^{N_{2}}\
e(\eta_{i},g)^{H(I_{i_{j}})} \cdot 
   \\ e(\sigma,g)^{-c_{u}}\cdot  e(\vartheta,g)^{\alpha'}\cdot
      e(\vartheta,\tilde {g})^{\alpha}  \wedge~ \big(Z_{l}h^{-c_{l}}=\\
g^{\gamma_{1}}\prod_{i=0}^{k-1}\tilde{h}_{i}^{w_{l_{i}}}
 \wedge Z_{l}\hat{g}_{1}^{-(d_{l}-q^{k})}=g^{\gamma_{1}}\prod_{i=0}^{k-1}\tilde{h}_{i}^{w'_{_{l_{i}}}}   \\
 \wedge (e(A_{w_{l_{i}}},\tilde{h})=e(h,h)^{t_{l}}\cdot e(A_{w_{l_{i}}},h)^{-w_{l_{i}}})_{j=0}^{k-1}~\wedge\\
      (e(A'_{w_{l_{i}}},\tilde{h})=e(h,h)^{t_{l}}\cdot e(A'_{w_{l_{i}}},h)^{-w'_{l_{i}}})_{j=0}^{k-1}\big)_{l=1}^{N_{1}}\\
   \wedge~ \left(e(B_{i_{j}},\tilde{\eta}_{i})=e(\eta,\eta)^{e_{i}}\cdot e(B_{i_{j}},\eta)^{H(I_{i_{j}})}\right)_{i=1}^{N_{2}}
\end{array}\right\}.
\end{equation*}}
 to extract $(x_{u}, c_{u},r_{u}, d,\alpha,\beta,\alpha',\beta', (a_{l}, (t_{l_{i}},t'_{l_{i}},w_{l_{i}},
$\\$w'_{l_{i}})_{i=0}^{k-1})_{l=1}^{N_{1}}, (e_{i}, H(I_{i_{j}}))_{i=1}^{N_{2}})$. If the extractor fails, $Sim_{1}$ returns $\perp$ to $\mathcal{E}$ to indicate failure; otherwise, $Sim_{1}$ runs $\mathcal{A}$ interacting with the honest ticket seller. The difference between {\bf Hybrid$_{\mathcal{E},Sim_{0}}$} and {\bf Hybrid$_{\mathcal{E},Sim_{1}}$} is 
\begin{equation*}
\left| {\bf Hybrid_{\mathcal{E},Sim_{0}}}-  {\bf Hybrid_{\mathcal{E},Sim_{1}}}\right|\leq \frac{q_{T}}{2^{\ell}}
\end{equation*}
where $q_{T}$ is the number of ticket issue queries.

\medskip
\noindent{\bf Game-2.} The simulator $Sim_{2}$ runs exactly as $Sim_{1}$ except that $Sim_{2}$ returns $\perp$ to $\mathcal{E}$ if one of the credentials $(c_{u},r_{u},\sigma_{U}=C\vartheta^{-\alpha})$ is not generated by the {\sf Join} algorithm. Actually, $(c_{u},r_{u},\sigma_{U}=C\vartheta^{-\alpha})$  is a forged BBS+ signature \cite{asm:ac2006} on $(x_{u},a_{u},r'_{u},((H(I_{i_{j}})_{A_{U}\models I_{i_{j}}} )_{i=1}^{N})$. For the case that multiple corrupted users pool their credentials, one of the pooled credentials must have a different $x_{u}$  when it was issued since only a single $x_{u}$ is extracted, hence is a forged credential. Due to the security of  the signature scheme \cite{asm:ac2006}, the difference between {\bf Hybrid$_{\mathcal{E},Sim_{2}}$} and {\bf Hybrid$_{\mathcal{E},Sim_{1}}$} is the following lemma.

\begin{claim}\label{c:1} We claim that
\begin{equation*}
\left|  {\bf Hybrid_{\mathcal{E},Sim_{2}}} - {\bf Hybrid_{\mathcal{E},Sim_{1}}} \right|\leq \frac{1}{q_{I}}Adv_{\mathcal{A}}^{q_{I}-SDH}(\ell),
\end{equation*}
where $q_{I}$ is the number of crendentail queries made by the adversary $\mathcal{A}$.
\end{claim}

\noindent{\bf Game-3.} The simulator $Sim_{3}$ runs exactly as $Sim_{2}$, except that $a_{l}\notin [c_{l},d_{l}]$. In this case, there exists at least one $w_{l_{i}}\notin [0,1,\cdots,q-1]$ or $w'_{l_{i}}\notin[0,1,\cdots,q-1]$. If $w_{l_{i}}\notin [0,1,\cdots,q-1]$ or $w'_{l_{i}}\notin[0,1,\cdots,q-1]$, we have $h_{w_{i}}=A_{w_{i}}^{\frac{1}{t_{i}}}=h^{\frac{1}{y+w_{i}}}$ or $h_{w'_{i}}=(A'_{w_{i}})^{\frac{1}{t'_{i}}}=h^{\frac{1}{y+w'_{i}}}$ is a forged BB signature \cite{bb:ss2004}. The difference between  ${\bf Hybrid_{\mathcal{E},Sim_{3}}}$ and $ {\bf Hybrid_{\mathcal{E},Sim_{2}}}$ is bounded by the following lemma.

\begin{claim}\label{c:2}
We claim that 
\begin{equation*}
\left|  {\bf Hybrid_{\mathcal{E},Sim_{3}}} - {\bf Hybrid_{\mathcal{E},Sim_{2}}} \right|\leq Adv_{\mathcal{A}}^{(q+1)-SDH}(\ell).
\end{equation*}
\end{claim}

\noindent{\bf Game-4.} The simulator $Sim_{4}$ runs exact as $Sim_{4}$ except that there exists at least an $I_{i_{j}}\notin \mathbb{S}_{i}$. If $I_{i_{j}}\notin \mathbb{S}_{i}$, so $B^{\frac{1}{e_{i}}}_{i_{j}}=\eta_{i_{j}}=\eta^{\frac{1}{\eta_{i}+H(I_{i_{j}})}}$ is a forged BB signature on $H(i_{j})$. 
The difference between  ${\bf Hybrid_{\mathcal{E},Sim_{4}}}$ and $ {\bf Hybrid_{\mathcal{E},Sim_{3}}}$ is bounded by the following lemma.
\begin{claim}\label{c:4}
We claim that 
\begin{equation*}
\left|  {\bf Hybrid_{\mathcal{E},Sim_{4}}} - {\bf Hybrid_{\mathcal{E},Sim_{3}}} \right|\leq Adv_{\mathcal{A}}^{(\varsigma+1)-SDH}(\ell).
\end{equation*}
\end{claim}

\noindent{\bf Game-5.} The simulator $Sim_{5}$ runs exactly as $Sim_{4}$ except that $Sim_{5}$ returns  tickets to $\mathcal{E}$.  At the first ticket issuing query dictated by $\mathcal{E}$,  $Sim_{5}$ runs the simulated proof of knowledge 
\begin{equation*}
\mbox{PoK}\left\{\begin{array}{ll} (c_{s},r_{s},\sigma_{S},z,v):
Q=\sigma_{S}\vartheta^{z} \wedge Z=g^{z}\vartheta^{v} \\
\wedge Z^{c_{s}}=g^{z'}\vartheta^{v'}\wedge
\frac{e(Q,\tilde{g})}{e(g_{0},g)\cdot e(g_{1},g)^{H(VP_{S})}}=\\
e(\rho,g)^{x_{s}}\cdot e(\mathfrak{g},g)^{r_{s}}
\cdot e(Q,g)^{-c_{s}}\cdot e(\vartheta,g)^{c_{s}z}\cdot\\
 e(\vartheta,\tilde{g})^{z}
\end{array}
\right\}.
\end{equation*}
The  tickets $(d_{u},\omega_{u},T_{u})$ for each ticket issuing query is computed by using the signing oracle in \cite{asm:ac2006}. The following lemma is used to bound the difference between ${\bf Hybrid_{\mathcal{E},Sim_{5}}} $ and $ {\bf Hybrid_{\mathcal{E},Sim_{4}}} $.
\begin{claim}\label{c:5}
We claim that 
\begin{equation*}
\left|{\bf Hybrid_{\mathcal{E},Sim_{5}}} - {\bf Hybrid_{\mathcal{E},Sim_{4}}}\right|\leq \frac{1}{q_{T}}Adv_{\mathcal{A}}^{q_{T}-SDH}(\ell).
\end{equation*}
where $q_{T}$ is the number of ticket issuing queries made by $\mathcal{A}$.
\end{claim}

\noindent{\bf Game-6.} The simulator $Sim_{6}$ runs exactly as $Sim_{5}$ except that $Sim_{6}$ runs the extractor of the proof of knowledge 
\begin{equation*}
\mbox{PoK}\left\{
\begin{array}{ll}
(x_{u},d_{u},s_{u},\omega_{u},\pi,\lambda,\pi',\lambda'): 
D=g^{s_{u}} \wedge \\
E=\xi^{x_{u}}H'(ID_{V})^{rs_{u}}
\wedge F=T_{U}\vartheta^{\pi}\wedge \\
 J=g^{\pi}\vartheta^{\lambda} \wedge J^{\omega_{u}}=g^{\pi'}\vartheta^{\lambda'}
\wedge 
  \frac{e(F,Y_{S})}{e(g_{0},\rho)e(g_{3},\rho)^{\psi_{u}}}\\
   =e(\xi,\rho)^{x_{u}}\cdot e(g_{1},\rho)^{d_{u}}\cdot e(g_{2},\rho)^{s_{u}}\cdot 
   \\
  e(F,\rho)^{-\omega_{u}}\cdot e(\vartheta,\rho)^{\pi'}\cdot e(\vartheta,\rho)^{\pi}\\
\end{array}
\right\}. 
\end{equation*}
to exact from $\mathcal{A}$ the witness  $(x_{u},d_{u},s,\omega_{u},\pi,\lambda,\pi',\lambda')$. If the extraction fails, $Sim_{7}$ returns $\perp$ to $\mathcal{E}$; otherwise, it continue to run $\mathcal{A}$ interacting with the honest verifier {\sf V}. The difference between     ${\bf Hybrid_{\mathcal{E},Sim_{6}}} $ and  ${\bf Hybrid_{\mathcal{E},Sim_{5}}}$ is 
\begin{equation*}
\left|{\bf Hybrid_{\mathcal{E},Sim_{6}}} - {\bf Hybrid_{\mathcal{E},Sim_{5}}}\right|\leq \frac{q_{V}}{2^{\ell}}
\end{equation*}
where $q_{V}$ is the number of ticket validation queries.
\medskip

\noindent{\bf Game-7.} The simulator $Sim_{7}$ runs exactly as $Sim_{6}$ except that $Sim_{7}$ returns $\perp$ to $\mathcal{E}$ if at least one of the extracted $(x'_{u},d'_{u},s',\omega'_{u},T'_{U},\pi,\lambda,$ $\pi',\lambda')$ was not generated by the Ticket Issuing algorithm. Actually, $(\omega'_{u},T'_{U})$ is a signature on $(x_{u},d'_{u},s')$. For multiple users case, one of the pooled tickets must have a different $x'_{u}$ than when it was issue since only one $x_{u}$ is extracted, hence $(\omega'_{u},T'_{U})$ is a forged signature on $(x'_{u},d'_{u},s')$. The following lemma is used to bound the difference between ${\bf Hybrid_{\mathcal{E},Sim_{8}}}$ and  ${\bf Hybrid_{\mathcal{E},Sim_{7}}}$.
\begin{claim}\label{c:6}
We claim that
\begin{equation*}
\left|{\bf Hybrid_{\mathcal{E},Sim_{7}}} - {\bf Hybrid_{\mathcal{E},Sim_{6}}}\right|\leq  \frac{1}{q_{V}}Adv_{\mathcal{A}}^{q_{V}-SDH}.
\end{equation*}
\end{claim}

\noindent{\bf Game-8.} Now, based on the real-world adversary $\mathcal{A}$, we construct an ideal-word adversary $\mathcal{A}'$. $\mathcal{A}'$ runs $\mathcal{A}$ to obtain $params$ and $Y_{S}$. After receiving a ticket issue query $(ticket\_issuing,Ps_{U'},\mathbb{P}_{U'},serv,Price,VP)$ from {\sf TP}, $\mathcal{A}'$ runs $\mathcal{A}$ to returns a simulated proof of the knowledge:
\begin{equation*}
\mbox{PoK}\left\{\begin{array}{ll} (c_{s},r_{s},\sigma_{S},z,v):
 Z=g^{z}\vartheta^{v} \wedge Z^{c_{s}}=g^{z'}\vartheta^{v'}\\
 \wedge \frac{e(Q,\tilde{g})}{e(g_{0},g)\cdot e(g_{1},g)^{H(VP_{S})}} =
e(\rho,g)^{x_{s}} 
\cdot e(\mathfrak{g},g)^{r_{s}} \cdot\\ e(Q,g)^{-c_{s}}\cdot e(\vartheta,g)^{c_{s}z}\cdot
 e(\vartheta,\tilde{g})^{z}
\end{array}
\right\}.
\end{equation*}
After  having extracted $(x_{u},c_{u},r_{u}, d_{u},\alpha,\beta,\alpha',\beta',t_{1},t_{2},$ $\cdots,t_{k-1},t'_{0}, t'_{1},\cdots,t'_{k-1}, $ $((e_{i}, H(I_{i_{j}}))_{A_{U}\models I_{i_{j}}})_{i=1}^{N})$ from $\mathcal{A}$, $\mathcal{A}'$ queries {\sf TP} to obtain a credential $(c_{u},r_{u},\sigma_{U})$ for ${\sf U}'$. Next, $\mathcal{A}'$ runs $\mathcal{A}$ to generate a BBS+ signature  \cite{asm:ac2006} $(s_{u},\omega_{u},T_{u})$ on $(x_{u},d_{u})$. If the signature can be generated correctly, {\sf TP}   returns $\mathcal{A}'$ with a bit $\hat{\nu}=1$; otherwise $\hat{\nu}=0$ is returned. After receiving a ticket validation query $(ticket\_$ $validating,T_{U'},Ps_{U'},\mathbb{P}_{U'},Serv,Price,VP_{T})$ from ${\sf V}'$,  $\mathcal{A}'$ runs $\mathcal{A}$ to execute the proof of knowledge 
\begin{equation*}
\mbox{PoK}\left\{
\begin{array}{ll}
(x_{u},d_{u},s_{u},\omega_{u},\pi,\lambda,\pi',\lambda'): 
D=g^{s_{u}}\wedge\\
Ps_{U}=\xi^{x_{u}}g_{1}^{d_{u}}\wedge E=\xi^{x_{u}}H'(ID_{V})^{rs_{u}}   \\
 J=g^{\pi}\vartheta^{\lambda} \wedge J^{\omega_{u}}=g^{\pi'}\vartheta^{\lambda'} \wedge \\
  \frac{e(F,Y_{S})}{e(g_{0},\rho)\cdot e(Ps_{U},\rho)\cdot e(g_{3},\rho)^{\psi_{u}}} 
  =e(g_{2},\rho)^{s_{u}}\cdot 
   \\
 e(F,\rho)^{-\omega_{u}}\cdot e(\vartheta,\rho)^{\pi'}\cdot e(\vartheta,\rho)^{\pi}\\
\end{array}
\right\}. 
\end{equation*}
If the proof is correct, {\sf TP} returns $\mathcal{A}'$  with a bit
$\bar{\nu}=1$  and adds $(T_{U'},Ps_{U'},\mathbb{P}_{U'},Serv,Price,VP_{T})$
into the ticket validation list; otherwise, $\bar{\nu}=0$ is returned. After
receiving a double spend detecting query $(double\_spend\_detecting,T_{U'})$
from ${\sf V}' $, ${\sf TP}$ checks where $T_{U'}$ is in the ticket validation
list. If it is, {\sf TP} returns ${\sf V}'$ with  a bit $\check{\nu}=1$;
otherwise, $\check{\nu}=0$ is returned. $\mathcal{A}'$ provides $\mathcal{A}$
with the same environment as $Sim_{7}$ did, hence we have
${\bf Hybrid}_{\mathcal{E},Sim_{8}} = {\bf Hybrid}_{\mathcal{E},Sim_{7}}$.

Therefore, we have 
\begin{equation*}
\begin{split}
&\left|{\bf Real_{\mathcal{E},\mathcal{A}}} - {\bf Ideal_{\mathcal{E},\mathcal{A}'}}\right|=
 \frac{q_{T}}{2^{\ell}}+\frac{q_{v}}{2^{\ell}}+\frac{1}{q_{I}}Adv_{\mathcal{A}}^{q_{I}-SDH}(\ell)\\
& +Adv_{\mathcal{A}}^{(q+1)-SDH}(\ell)+ Adv_{\mathcal{A}}^{(\varsigma+1)-SDH}(\ell)
+\\
&\frac{1}{q_{T}}Adv_{\mathcal{A}}^{q_{T}-SDH}(\ell)+\frac{1}{q_{V}}Adv_{\mathcal{A}}^{q_{V}-SDH}(\ell).
\end{split}
\end{equation*}
\end{proof}

\noindent{\bf Proof of Claim \ref{c:1}.} This claim is proven by constructing an algorithm $\mathcal{B}$ that can break the unforgeability under the adaptively chosen message attacks of BBS+ signature \cite{asm:ac2006}. From the security proof presented in \cite{asm:ac2006}, there exist a polynomial-time algorithm $\mathcal{B}$ that can break the $q_{I}$-SDH assumption with non-negligible advantage. 

Suppose that the adversary $\mathcal{A}$ can distinguish {\bf Game-1} and {\bf Game-2}. Given $(\rho,\rho^{x_{s}},\rho^{x_{s}^{2}},\cdots,\rho^{x_{s}^{q_{I}}})$. $\mathcal{B}$ aims to outpput $(c,\rho^{\frac{1}{x_{s}+c}})$ where $c\in\mathbb{Z}_{p}$ and $c\neq -x_{s}$. Let $Y_{S}=\rho^{x_{s}}$.

$\mathcal{B}$ sends $q_{I}$ messages $(m_{1},m_{2},\cdots,m_{q_{I}})$ to the challenger $\mathcal{C}$, and obtains $(\tilde{\sigma}_{1},\tilde{\sigma_{2}},\cdots,\tilde{\sigma}_{q_{I}})$ where $\tilde{\sigma}_{i}=\rho^{\frac{1}{x_{s}+m_{i}}}$ and $e(\tilde{\sigma}_{i},Y_{S}\rho^{m_{i}})=e(\rho,\rho)$ for $i=1,2,\cdots,q_{I}$. $\mathcal{B}$ selects $\tilde{\alpha},\tilde{\beta},\tilde{\gamma},\tilde{a},\tilde{b},\tilde{c},\tilde{d}\stackrel{R}{\leftarrow}\mathbb{Z}_{p}$, and computes 
$g_{1}=((Y_{S}\rho^{\tilde{\alpha}})^{\tilde{\gamma}}\rho^{-1})^{\frac{1}{\tilde{\beta}}}=\rho^{\frac{(x_{s}+\tilde{\alpha})\tilde{\gamma}-1}{\tilde{\beta}}}$, $g_{0}=\rho^{\tilde{a}}$, $g_{2}=g_{1}^{\tilde{b}}$, $g_{3}=g_{1}^{\tilde{d}}$ and $\xi=g_{1}^{\tilde{c}}$. $\mathcal{B}$ sends $\mathcal{A}$  $(\rho,Y_{S},g_{0},\xi,g_{1},g_{2},$ $g_{3})$.

For $q_{I}$ ticket issuing queries, $\mathcal{B}$ selects one and referred as query $\tilde{Q}=(Ps_{U},\mathbb{P}_{U},Price,Serv,VP_{U})$ where $Ps_{U}=Y=\xi^{x_{u}}g_{1}^{d_{u}}$. For $q_{I}-1$ queries other than query $\tilde{Q}$, $\mathcal{B}$ responses using the $q_{I}-1$ pairs $(m_{i},\sigma_{i})$ as follows.

Suppose that $\mathcal{A}$ queries a ticket on $(Ps_{U},\mathbb{P}_{U},Price,$ $Serv,VP_{T})$, since $\mathcal{B}$  extracts  the knowledge of  $(x_{u}, c_{u},r_{u}, d,\alpha,\beta,\alpha',\beta',(a_{l}, (t_{l_{i}},t'_{l_{i}},w_{l_{i}},
w'_{l_{i}})_{i=0}^{k-1})_{l=1}^{N_{1}},$ $ (e_{i}, H(I_{i_{j}}))_{i=1}^{N_{2}})$,     $\mathcal{B}$ selects $d',s_{u}\stackrel{R}{\leftarrow}\mathbb{Z}_{p}$, and computes $\psi_{u}=H(\mathbb{P}_{U}||Price||Services||VP_{T})$ and  $t=x_{u}\tilde{c}+(d'+d)+\tilde{b}s_{u}+\tilde{d}\psi_{u}$ and 
\begin{equation*}
\begin{split}
T_{U}&=(g_{0}Yg_{1}^{d'}g_{2}^{s_{u}}g_{3}^{\psi_{u}})^{\frac{1}{x_{s}+m_{i}}}=(g_{0}g_{1}^{t})^{\frac{1}{x_{s}+m_{i}}}=\sigma_{i}^{\tilde{a}}g_{1}^{\frac{t}{x_{s}+m_{i}}}\\
&=\sigma_{i}^{\tilde{a}} \rho^{\frac{t((x_{s}+\tilde{\alpha})\tilde{\gamma}-1)}{\tilde{\beta}(x_{s}+m_{i})}}=\sigma_{i}^{\tilde{a}}(\rho^{\frac{(x_{s}+\tilde{\alpha})\tilde{\gamma}-1}{(x_{s}+m_{i})}})^{\frac{t}{\tilde{\beta}}}\\
&=\sigma_{i}^{\tilde{a}}(\rho^{\frac{(x_{s}+m_{i}-m_{i}+\tilde{\alpha})\tilde{\gamma}-1}{(x_{s}+m_{i})}})^{\frac{t}{\tilde{\beta}}}=\sigma_{i}^{\tilde{a}}\rho^{\frac{\tilde{t\gamma}}{\tilde{\beta}}} \sigma_{i}^{\frac{t((\tilde{\alpha}-m_{i})\tilde{\gamma}-1)}{\tilde{\beta}}}
\end{split}
\end{equation*}
where $m_{i}\in\{m_{1},m_{2},\cdots,m_{q_{I}}\}$.

For the query $\tilde{Q}$  where $\tilde{Y}=\xi^{\tilde{x}}g_{1}^{\tilde{d}}$, $\mathcal{B}$ computes $\tilde{\psi}=H(\mathbb{P}_{\tilde{U}}||Price||Serv||VP_{T})$ and selects $\tilde{d}',\tilde{s}\in\mathbb{Z}_{p}$ such that $\tilde{d}'+\tilde{d}+\tilde{c}\tilde{x}+\tilde{b}\tilde{s}+\tilde{d}\tilde{\psi}=\tilde{a}\tilde{\beta}$ and computes 
\begin{equation*}
\begin{split}
\tilde{T}&=(g_{0}\tilde{Y}g_{1}^{\tilde{d}'}g_{2}^{\tilde{s}}g_{3}^{\tilde{\psi}})^{\frac{1}{x_{s}+\tilde{\alpha}}}
=e(g_{0}g_{1}^{\tilde{c}\tilde{x}+\tilde{d}+\tilde{d}'+\tilde{b}\tilde{s}+\tilde{d}\tilde{\psi}})^{\frac{1}{x_{s}+\tilde{\alpha}}}\\
&=(\rho^{\tilde{a}}\rho^{\frac{((x_{s}+\tilde{\alpha})\tilde{\gamma}-1)(\tilde{c}\tilde{x}+\tilde{d}+\tilde{d}'+\tilde{b}\tilde{s}+\tilde{d}\tilde{\psi})}{\tilde{\beta}}})^{\frac{1}{x_{s}+\tilde{\alpha}}}\\
&=(\rho^{\frac{((x_{s}+\tilde{\alpha})\tilde{\gamma}-1)(\tilde{c}\tilde{x}+\tilde{d}+\tilde{d}'+\tilde{b}\tilde{s})+\tilde{a}\tilde{\beta}}{\tilde{\beta}}})^{\frac{1}{x_{s}+\tilde{\alpha}}}\\
&=(\rho^{\frac{((x_{s}+\tilde{\alpha})\tilde{\gamma}-1)\tilde{a}\tilde{\beta}+\tilde{a}\tilde{\beta}}{\tilde{\beta}}})^{\frac{1}{x_{s}+\tilde{\alpha}}}=\rho^{\tilde{\gamma}\tilde{a}}.
\end{split}
\end{equation*}
$\mathcal{B}$ repondes $\mathcal{A}$ with $(\tilde{a}\tilde{\beta}-\tilde{d}-\tilde{c}\tilde{x}-\tilde{b}\tilde{s}-\tilde{d}\tilde{\psi},\tilde{s},\tilde{\psi},\alpha,\tilde{T})$ where $\tilde{\psi}=H(\mathbb{P}_{\tilde{U}}||Price||Serv||VP_{T})$.
  Finally, $\mathcal{A}$ output a forged tickets $(\tilde{d}^{*},s^{*},\psi^{*},\omega^{*},T^{*})$ for $(Y^{*},\mathbb{P}_{U^{*}}, $ $price, Serv, VP)$ where $\psi^{*}=H(\mathbb{P}_{U^{*}}||$ $Price||Serv||VP_{T})$, $Y^{*}=\xi^{x^{*}}g_{1}^{d^{*}}$ and $T^{*}=(g_{0}Y^{*}g_{1}^{\tilde{d}^{*}}g_{2}^{s*}g_{3}^{\psi^{*}})^{\frac{1}{x_{s}+\omega^{*}}}$. $\mathcal{B}$ runs $\mathcal{A}$ to extract $(x^{*},d^{*})$ from the proof of $\prod_{U^{*}}^{2}$.  We consider the following  cases.

  \noindent{\em Case-I}$~(\omega^{*}\notin(m_{1},m_{2},\cdots,m_{q_{I}},\tilde{\alpha}))$: Let $t^{*}=\tilde{c}x^{*}+d^{*}+\tilde{d}^{*}+\tilde{b}s^{*}+\tilde{d}\psi^{*}$.

$T^{*}=(g_{0}Y^{*}g_{1}^{\tilde{d}^{*}}g_{2}^{s*}g_{3}^{\psi^{*}})^{\frac{1}{x_{s}+\omega^{*}}}=\\
(g_{0}g_{1}^{\tilde{c}x^{*}+d^{*}+\tilde{d}^{*}+\tilde{b}s^{*}+\tilde{d}\psi^{*}})^{\frac{1}{x_{s}+\omega^{*}}}
=(\rho^{\tilde{a}}\rho^{\frac{t^{*}((x_{s}+\tilde{\alpha})\tilde{\gamma}-1)}{\tilde{\beta}}})^{\frac{1}{x_{s}+\omega^{*}}}\\
=\rho^{\frac{\tilde{a}}{x_{s}+\omega^{*}}}\rho^{\frac{t^{*}((x_{s}+\tilde{\alpha})\tilde{\gamma}-1)}{\tilde{\beta}(x_{s}+\omega^{*})}}
=\rho^{\frac{\tilde{a}}{x_{s}+\omega^{*}}}\rho^{\frac{t^{*}((x_{s}+\omega^{*}-\omega^{*}+\tilde{\alpha})\tilde{\gamma}-1)}{\tilde{\beta}(x_{s}+\omega^{*})}}\\
=\rho^{\frac{\tilde{a}}{x_{s}+\omega^{*}}}\rho^{\frac{t^{*}\tilde{\gamma}}{\tilde{\beta}}}\rho^{\frac{t^{*}(\tilde{\alpha}-\omega^{*})\tilde{\gamma}}{\tilde{\beta}(x_{s}+\omega^{*})}}\rho^{\frac{-t^{*}}{\tilde{\beta}(x_{s}+\omega^{*})}}.
$

We have
 $  T^{*}\rho^{\frac{-t^{*}\tilde{\gamma}}{\tilde{\beta}}}=\rho^{\frac{\tilde{a}\tilde{\beta}+t^{*}(\tilde{\alpha}-\omega^{*})\tilde{\gamma}-t^{*}}{\tilde{\beta}(x_{s}+\omega^{*)}}}$ 
  and 
$
  \rho^{\frac{1}{x_{s}+\omega^{*}}}=$ $(T^{*}\rho^{\frac{-t^{*}\tilde{\gamma}}{\tilde{\beta}}})^{\frac{\tilde{\beta}}{\tilde{a}\tilde{\beta}+t^{*}(\tilde{\alpha}-\omega^{*})\tilde{\gamma}-t^{*}}}
$. $\mathcal{B}$ outputs $(\omega^{*},(T^{*}\rho^{\frac{-t^{*}\tilde{\gamma}}{\tilde{\beta}}})^{\frac{\tilde{\beta}}{\tilde{a}\tilde{\beta}+t^{*}(\alpha^{*}-\omega^{*})\tilde{\gamma}-t^{*}}})$.

  \noindent{\em Case-II}~$(\omega^{*}=m_{i}~\mbox{and}~T^{*}=T_{i})$ or $(\omega^{*}=\tilde{\alpha}~\mbox{and}~T^{*}=\tilde{T})$: These happen with negligible probability except that $\mathcal{A}$ can solves the related discrete logarithms amongst off $g_{0},g_{1},g_{2},g_{3}$ and $\xi$.

  \noindent{\em Case-III}~ $(\omega^{*}\in(m_{1},m_{2},\cdots,m_{q_{I}},\tilde{\alpha}))~\mbox{and}~T^{*}\neq T_{i}~\mbox{or}~T^{*}\neq \tilde{T}$: If it is, $\omega^{*}=\tilde{\alpha}$ with the probability $\frac{1}{q_{I}}$. Let $t^{*}=\tilde{c}x^{*}+d^{*}+\tilde{d}^{*}+\tilde{b}s^{*}+\tilde{d}\psi^{*}$. We have
  $\rho^{\frac{1}{x_{s}+\omega^{*}}}=(T^{*}\rho^{\frac{-t^{*}\tilde{\gamma}}{\tilde{\beta}}})^{\frac{\tilde{\beta}}{\tilde{a}\tilde{\beta}-t^{*}}}$. 
    $\mathcal{B}$ outputs $(\omega^{*},(T^{*}\rho^{\frac{-t^{*}\tilde{\gamma}}{\tilde{\beta}}})^{\frac{\tilde{\beta}}{\tilde{a}\tilde{\beta}-t^{*}}})$.
  \medskip

  Therefore, 
$ \left|{\bf Hybrid}_{\mathcal{E},Sim_{2}}(\ell)-{\bf Hybrid}_{\mathcal{E},Sim_{1}}(\ell)\right|\leq\\
   \frac{Adv_{\mathcal{A}}^{q_{I}-SDH}(\ell)}{q_{I}}.$
\medskip

\noindent{\bf Proof of Claim \ref{c:2}. } This claim is proven by constructing an algorithm $\mathcal{B}$ that can break the unforgeability under the weak chosen message attacks of BB signature \cite{bb:ss2004}. By the security proof given in \cite{bb:ss2004}, there exists an polynomial-time algorithm $\mathcal{B}$ that can break the $(q+1)$-SDH assumption with non-negligible advantage.
 
 Suppose that an adversary $\mathcal{A}$ can distinguish {\bf Game-2} and {\bf Game-3}.  Given $(h,h^{y},h^{y^{2}},\cdots,h^{y^{q+1}})$, $\mathcal{B}$ aims to output $(c,h^{\frac{1}{y}})$ where $c\in\mathbb{Z}_{p}$ and $c\neq -y$. Receiving $q$ messages $\{0,1,\cdots,q-1\}$ from $\mathcal{A}$, $\mathcal{B}$ computes $f(y)=\prod_{j=0}^{q}(y+j)=\sum_{j=0}^{q}\pi_{j}y^{j}$, $yf(y)=\sum_{j=1}^{q+1}\phi_{j}y^{j}$ and $f_{i}(y)=\frac{f(y)}{y+i}=\sum_{j=0}^{q-1}\varpi_{j} y^{j}$ where $\pi_{0},\cdots,\phi_{q},\phi_{1},\cdots,\phi_{q+1},\varpi_{0},\cdots,$ $\varpi_{q-1}\in\mathbb{Z}_{p}$. $\mathcal{B}$ computes $\hat{h}=h^{f(y)}=\prod_{j=0}^{q}(h^{y^{j}})^{\pi_{i}}$, $\tilde{h}=\hat{h}^{y}=h^{yf(y)}=\prod_{j=0}^{q}(h^{y^{j+1}})^{\pi_{j}}$ and  ${h}_{i}=\hat{h}^{\frac{1}{y+i}}=h^{\frac{f(y)}{y+i}}\prod_{j=0}^{q-1}(h^{y^{j}})^{\varpi_{i}}$ for $i=0,2,\cdots,q-1$. $\mathcal{B}$ sends $(\hat{h},\tilde{h},{h}_{1},{h}_{2},\cdots,{h}_{q})$ to $\mathcal{A}$. Since $\mathcal{B}$ extracts $(x_{u},a_{u},c_{u},r_{u},d_{u},\alpha,\beta,\alpha',\beta',(t_{i},t'_{i},\omega_{i},\omega'_{i})_{i=0}^{k-1},$ $ ((e_{i},H(I_{i_{j}}))_{A_{U}\in I_{i_{j}}})_{i=1}^{N})$ with $e(A_{w_{i}},\tilde{h})=e(h,h)^{t_{i}}\cdot e(A_{w_{i}},h)^{-w_{i}}$ and $e(A'_{w_{i}},\tilde{h})=e(h,h)^{t_{i}}\cdot e(A'_{w_{i}},h)^{-w'_{i}}$. Hence, $h_{w_{i}}=(A_{w_{i}})^{\frac{1}{t_{i}}}=\hat{h}^{\frac{1}{y+w_{i}}}$ or $h_{w_{i}}=(A'_{w_{i}})^{\frac{1}{t'_{i}}}=\hat{h}^{\frac{1}{y+w'_{i}}}$. When $w_{i}\notin\{0,1,\cdots,k-1\}$, let $f(y)=c(x)\cdot(x+w_{i})+\gamma$, where  $\gamma\neq 0$ and $c(x)=\sum_{j=0}^{q-1}\varrho_{j}y^{j}$ is a $(q-1)$-degree polynomial. We have  
$
  h_{w_{i}}=\hat{h}^{\frac{1}{y+w_{i}}}=h^{\frac{f(y)}{y+w_{i}}}=h^{\frac{\gamma+c(x)\cdot(x+w_{i})}{y+w_{i}}}=h^{\frac{\gamma}{y+w_{i}}}\cdot\prod_{j=0}^{q-1}(h^{y^{j}})^{\varrho_{j}}
$
   and  $h^{\frac{1}{y+w_{i}}}=(h_{w_{i}}\prod_{j=0}^{q-1}(h^{y^{j}})^{-\psi_{j}})^{\frac{1}{\gamma}}=((A_{w_{i}})^{\frac{1}{t_{i}}}\prod_{j=0}^{q-1}(h^{y^{j}})^{-\varrho_{j}})^{\frac{1}{\gamma}}.$ Finally,  $\mathcal{B}$ outputs $(w_{i},h^{\frac{1}{y+w_{i}}})$. Therefore, 
 \begin{equation*}
 \left|{\bf Hybrid}_{\mathcal{E},Sim_{3}}-{\bf Hybrid}_{\mathcal{E},Sim_{2}}\right|\leq Adv_{\mathcal{A}}^{(q+1)-SDH}(\ell).
 \end{equation*}

 \noindent{\bf Proofs of Claim  \ref {c:4}, \ref{c:5}, \ref{c:6}. }  The proof of Claim  \ref {c:4} is similar to the proof of Claim \ref{c:2}.
The proofs of Claim \ref{c:5} and \ref{c:6} is similar as the proof of Claim \ref{c:1}.

\subsection{The Details of Zero-Knowledge Proof}\label{app_zkp}
The details of zero-knowledge proofs used in our PPETS-FGP are described by using the Fiat-Shamir heuristic \cite{fs:fs1986} as follows.
\medskip

\noindent{\bf The Detail of $\prod_{S}^{1}$:}  

\noindent  {\sf S} select $t_{s}\stackrel{R}{\leftarrow}\mathbb{Z}_{p}$ and $M_{S}^{1}\stackrel{R}{\leftarrow}\mathbb{G}$, and computes $T_{S}=\rho^{t_{s}}$, $c=H(M_{S}^{1}||Y_{S}||T_{S})$ and $s=t_{s}-cx_{s}$. {\sf S}  sends $(c,s,M_{S}^{1},Y_{S})$ to  {\sf CA}.

\noindent {\sf CA} verifies $c\stackrel{?}{=}H(M_{S}^{1}||Y_{S}||\rho^{s}Y_{S}^{c})$.
\medskip

\noindent{\bf The Detail of $\prod_{U}^{1}$:}

\noindent {\sf U} select $\tilde{x},\tilde{r}\stackrel{R}{\leftarrow}\mathbb{Z}_{p}$ and $M_{U}^{1}\stackrel{R}{\leftarrow}\mathbb{G}$, and computes $Y'_{U}=\xi^{\tilde{x}}$, $R'=g^{\tilde{r}}$, $c_{1}=H(M_{U}^{1}||Y_{U}||Y'_{U})$, $c_{2}=H(M_{U}^{1}||R||R')$, $s_{1}=\tilde{x}-c_{1}x_{u}$ and $s_{2}=\tilde{r}-c_{2}r$. {\sf U} sends $(M_{U}^{1},Y_{U},R,c_{1},c_{2},s_{1},s_{2})$ to {\sf CA}.

\noindent {\sf CA} verifies $c_{1}\stackrel{?}{=}H(M_{U}^{1}||Y_{U}||\xi^{s_{1}}Y_{U}^{c_{1}})$ and $c_{2}=H(M_{U}^{1}||R||g^{s_{2}}R^{c_{2}})$.
\medskip

\noindent{\bf The Detail of $\prod_{S}^{2}$:}

\noindent {\sf S} selects $z,v,\tilde{z},\tilde{v},\hat{z},\hat{v},\tilde{x}_{s},\tilde{v}_{s},\tilde{c}_{s}\stackrel{R}{\leftarrow}\mathbb{Z}_{p}$ and $M_{S}^{2}\stackrel{R}{\leftarrow}\mathbb{G}$, and computes 

\begin{equation*}
\begin{split}
&Q=\sigma_{S}\vartheta^{z},~Z=g^{z}\vartheta^{v},~ \Gamma=g^{zc_{s}}\vartheta^{cc_{s}}=g^{z'}\vartheta^{v'}, ~Z'=g^{\tilde{z}}\vartheta^{\tilde{v}},\\
& \Gamma'=g^{\hat{z}}\vartheta^{\hat{v}},~\Omega=\frac{e(Q,\tilde{g})}{e(g_{0},g)\cdot e(g_{1},g)^{H(VP_{S})}},~\Omega'=e(\rho,g)^{\tilde{x}_{s}}\cdot   \\
& e(\mathfrak{g},g)^{\tilde{v}_{s}}\cdot e(Q,g)^{-\tilde{c}_{s}}\cdot e(\vartheta,g)^{\hat{z}}\cdot e(\vartheta,\tilde{g})^{\tilde{z}},\\
\end{split}
\end{equation*}
\begin{equation*}
\begin{split}
& \tilde{c}_{1}=H(M_{S}^{2}||Z||Z'),
 \tilde{s}_{1}=\tilde{z}-\tilde{c}_{1}z, \tilde{s}_{2}=\tilde{v}-\tilde{c}_{1}v,\\
& \tilde{c}_{2}=H(M_{S}^{2}||\Gamma||\Gamma'), \hat{s}_{1}=\hat{z}-\tilde{c}_{2}z',\hat{s}_{2}=\hat{v}-\tilde{c}_{2}v',\\
& \tilde{c}_{3}=H(M_{S}^{2}||\Omega||\Omega'), \tilde{r}_{1}=\tilde{x}_{s}-\tilde{c}_{3}x_{s}, \tilde{r}_{2}=\tilde{v}_{s}-\tilde{c}_{3}r_{s},\\
& \tilde{r}_{3}=\tilde{c}_{s}-\tilde{c}_{3}c_{s},
 \tilde{r}_{4}=\hat{z}-\tilde{c}_{3}z',\tilde{r}_{5}=\tilde{z}-\tilde{c}_{3}z.
\end{split}
\end{equation*}

{\sf S} sends $(M_{S}^{2},Q,Z,\Gamma,\Omega,\tilde{c}_{1},\tilde{s}_{1},\tilde{s}_{2},\tilde{c}_{2},\hat{s}_{1},\hat{s}_{2},\tilde{c}_{3},\tilde{r}_{1},\tilde{r}_{2},$ $\tilde{r}_{3},\tilde{r}_{4},\tilde{r}_{5},VP_{S})$ to {\sf U}.

{\sf U} verifies: $ \Omega\stackrel{?}{=}\frac{e(Q,\tilde{g})}{e(g_{0},g)\cdot e(g_{1},g)^{H(VP_{S})}};$
\begin{equation*}
\begin{split}
&\tilde{c}_{1}\stackrel{?}{=}H(M_{S}^{2}||Z||g^{\tilde{s}_{1}}\vartheta^{\tilde{s}_{2}}Z^{\tilde{c}_{1}}); \tilde{c}_{2}\stackrel{?}{=}H(M_{S}^{2}||\Gamma||g^{\hat{s}_{1}}\vartheta^{\hat{s}_{2}}\Gamma^{\tilde{c}_{2}});\\
&\tilde{c}_{3}\stackrel{?}{=}H(M_{S}^{2}||\Omega||e(\rho,g)^{\tilde{r}_{1}}\cdot e(\mathfrak{g},g)^{\tilde{r}_{2}}\cdot e(Q,g)^{-\tilde{r}_{3}}\cdot e(\vartheta,g)^{\tilde{r}_{4}}\cdot\\
& e(\vartheta,\tilde{g})^{\tilde{r_{5}}}\cdot\Omega^{\tilde{c}_{3}}).
\end{split}
\end{equation*}

\noindent{\bf The Detail of $\prod_{U}^{2}$:}\\
\label{app:PI_U_2}
 {\sf U} selects $d,\alpha,\beta,(\gamma_{l},\tilde{\gamma}_{l},\tilde{a}_{l},(t_{l_{i}},t'_{l_{i}})_{i=0}^{k-1})_{l=1}^{N_{1}},(e_{j},\tilde{e}_{j},\tilde{e}'_{j},$ $\check{e}_{j})_{j=1}^{N_{2}},\tilde{d},$
 $\tilde{x}_{u},\tilde{r}_{u},\tilde{c}_{u},$  $\tilde{\alpha},\tilde{\beta},\tilde{c},((\tilde{t}_{l_{i}},\tilde{t}'_{l_{i}},\tilde{w}_{l_{i}},\tilde{w}'_{l_{i}})_{i=0}^{k-1})_{l=1}^{N_{1}}\stackrel{R}{\leftarrow}\mathbb{Z}_{p}$ and $M_{U}^{2}\stackrel{R}{\leftarrow}\mathbb{G}$, and computes
 \begin{equation*}
 \begin{split}
 & C=\sigma_{U}\vartheta^{\alpha}, D=g^{\alpha}\vartheta^{\beta}, \Phi=D^{c_{u}}=g^{\alpha c_{u}}\vartheta^{\beta c_{u}}=g^{\alpha'}\vartheta^{\beta'},\\
 &Y=\xi^{x_{u}}g_{1}^{d},\Big(Z_{l}=g^{\gamma_{l}}h^{a_{l}},{Z}'_{l}=g^{\tilde{\gamma}_{l}}h^{\tilde{a}_{l}},~\tilde{Z}_{l}=g^{\tilde{\gamma_{l}}}\prod_{i=0}^{k-1}\tilde{h}_{i}^{\tilde{w}_{l_{i}}},\\
 & \tilde{Z}'_{l}=g^{\tilde{\gamma_{l}}}\prod_{i=0}^{k-1}\tilde{h}_{i}^{\tilde{w}'_{l_{i}}},
 \big(A_{w_{l_{i}}}=h_{w_{l_{i}}}^{t_{l_{i}}},A'_{w_{l_{i}}}=h_{w'_{l_{i}}}^{t'_{{l_{i}}}},  V_{l_{i}}=\\
 &  e(h,h)^{t_{l_{i}}}\cdot e(A_{w_{l_{i}}},h)^{-w_{l_{i}}},\tilde{V}_{l_{i}}=e(h,h)^{\tilde{t}_{l_{i}}}\cdot e(A_{w_{l_{i}}},h)^{-\tilde{w}_{l_{i}}},\\
&  V'_{l_{i}}=e(h,h)^{t'_{l_{i}}}\cdot  e(A'_{w_{l_{i}}},h)^{-{w}'_{l_{i}}}, \tilde{V}'_{l_{i}}=e(h,h)^{\tilde{t}'_{l_{i}}}\cdot \\ &e(A'_{w_{l_{i}}},h)^{-\tilde{w}'_{l_{i}}})_{i=0}^{k-1}\Big)_{l=1}^{N_{1}},\tilde{D}=g^{\tilde{\alpha}}\vartheta^{\tilde{\beta}}, \\
& \tilde{\Phi}=D^{\tilde{c}}=g^{\tilde{\alpha}'}\vartheta^{\tilde{\beta}'} (\tilde{\alpha}'=\tilde{c}\alpha, \tilde{\beta}'=\tilde{c}\beta),\tilde{Y}=\xi^{\tilde{x}_{u}}g_{1}^{\tilde{d}},\\
&R=\frac{e(C,\tilde{g})}{e(g_{0},g)\cdot  e(g_{1},g)^{H(VP_{U})}},\\
& R'=e(\xi,g)^{\tilde{x}_{u}}\cdot e(\mathfrak{g},g)^{\tilde{r}_{u}}\cdot \prod_{l=1}^{N_{1}}e(\hat{g}_{l},g)^{\tilde{a}_{l}}\cdot\\
&\prod_{i=1}^{N_{2}}e(\eta_{i},g)^{\check{e}_{i}}
 \cdot e(C,g)^{-\tilde{c}_{u}}\cdot e(\vartheta ,g)^{\tilde{\alpha}'}\cdot e(\vartheta,\tilde{g})^{\tilde{\alpha}},\\
&\big(B_{i_{j}}=\eta_{i_{j}}^{e_{i}}, W_{i_{j}}=e(B_{i_{j}},\tilde{\eta}_{i}),\tilde{W}_{i_{j}}=e(\eta,\eta_{i})^{\tilde{e}_{i}}\cdot\\
& e(B_{i_{j}},\eta_{i})^{\tilde{e}'_{i}}\big)_{i=1}^{N_{2}},
\end{split}
\end{equation*}
\begin{equation*}
\begin{split}
&\bar{c}=H\big(M_{U}^{2}||Y||\tilde{Y}||D||\tilde{D}||\Phi||\tilde{\Phi}||C||R||R'||Z_{1}||\cdots||Z_{N_{1}}||\\
&{Z}'_{1}|| \cdots||{Z}'_{N_{1}}||B_{1_{1}}||\cdots||B_{1_{\zeta}}|| \cdots||B_{N_{2_{1}}}||\cdots|| B_{N_{2_{\zeta}}}||\\
& W_{1_{1}}||\cdots||W_{1_{\zeta}}||\cdots||W_{_{2_{1}}}||\cdots|| W_{N_{2_{\zeta}}}||\tilde{W}_{1_{1}}||\cdots||\tilde{W}_{1_{\zeta}}||\\
&\cdots||\tilde{W}_{N_{2_{1}}}||\cdots|| \tilde{W}_{N_{2_{\zeta}}}\big),~\bar{x}_{u}=\tilde{x}_{u}-\bar{c}x_{u}, ~\bar{d}=\tilde{d}-\bar{c}d,\\
&~ \bar{r}_{u}=\tilde{r}_{u}-\bar{c}r_{u}, ( \bar{\gamma}_{l}=\tilde{\gamma}_{l}-\bar{c}\gamma_{l},~\bar{a}_{l}=\tilde{a}_{l}-\bar{c}a_{l})_{l=1}^{N_{1}},\\
\end{split}
\end{equation*}
\begin{equation*}
\begin{split}
&(\hat{e}_{i}=\tilde{e}_{i}-\bar{c}e_{i}, ~\hat{e}'_{i}=\check{e}_{i}-\bar{c}H(I_{i_{j}}),~\hat{e}''_{i}=\check{e}_{i}+\bar{c}H(I_{i_{j}}))_{i=1}^{N_{2}}, \\
& \bar{c}_{u}=\tilde{c}_{u}-\bar{c}c_{u},~\bar{\alpha}=\tilde{\alpha}-\bar{c}\alpha,\bar{\beta}=\tilde{\beta}-\bar{c}\beta,\bar{\alpha}'=\tilde{\alpha}'-\bar{c}\alpha',\\
&\bar{\beta}'=\tilde{\beta}'-\bar{c}\beta', \big( \bar{e}_{l}=H(M_{U}^{2}||Z_{l}||Z'_{l}||\tilde{Z}_{l}||\tilde{Z}'_{l}), \check{\gamma}_{l}= \tilde{\gamma}_{l}-\\
& \bar{e}_{l}\gamma_{l}, \check{a}_{l}=\tilde{a}_{l}-\bar{e}_{l}(a_{l}-c_{l}),\check{a}_{l}'=\tilde{a}_{l}-\bar{e}_{l}(a_{l}-d_{l}+q^{k}),\\
&~(\bar{w}_{l_{i}}=\tilde{w}_{l_{i}}-\bar{e}_{l}w_{l_{i}},\bar{w}'_{l_{i}}=\tilde{w}'_{l_{i}}-\bar{e}_{l}w'_{l_{i}})_{i=0}^{k-1}\big)_{l=1}^{N_{1}},\\
&\big((\bar{d}_{l_{i}}=H(M_{U}^{2}||A_{w_{l_{i}}}||A'_{w_{l_{i}}}||V_{{l_{i}}}||V'_{{l_{i}}}||\tilde{V}_{{l_{i}}}||\tilde{V}'_{{l_{i}}}),  
\end{split}
\end{equation*}
\begin{equation*}
\begin{split}
&  \bar{t}_{l_{i}}=\tilde{t}_{l_{i}}-\bar{d_{l_{i}}}t_{l_{i}}, \bar{t}'_{l_{i}}=\tilde{t}'_{l_{i}}-\bar{d_{l_{i}}}t'_{l_{i}},\hat{w}_{l_{i}}=\tilde{w}_{l_{i}}-\bar{d}_{l_{i}}w_{l_{i}},\\
& \hat{w}'_{l_{i}}=\tilde{w}'_{l_{i}}-\bar{d}_{l_{i}}w'_{l_{i}})_{i=0}^{k-1}\big)_{l=1}^{N_{1}},\\
\end{split}
 \end{equation*}
{\sf U} sends {\sf S}:

\noindent $\big(C,D,\Phi,Y,R,(Z_{l},(\tilde{A}_{w_{l_{i}}},\tilde{A}_{w_{l_{i}}}',V_{l_{i}},\tilde{V}_{l_{i}},V'_{l_{i}})_{i=0}^{k-1})_{l=1}^{N_{1}},$ $({B}_{i_{j}},W_{i_{j}})_{i=1}^{N_{2}}, $ $\bar{c},\bar{x}_{u},\bar{d},\bar{r}_{u},\bar{c}_{u},\bar{\alpha},\bar{\beta},\bar{\alpha}',\bar{\beta}',$ $
(\bar{e}_{l},\bar{\gamma}_{l},\bar{a}_{l},\check{\gamma}_{l},$ $\check{a}_{l},\bar{a}'_{l})_{l=1}^{N_{1}},(\hat{e}_{i},\hat{e}'_{i})_{i=1}^{N_{2}}((\bar{w}_{l_{i}},\bar{w}'_{l_{i}},\hat{w}_{l_{i}},\hat{w}'_{l_{i}},\bar{d}_{l_{i}},\bar{t}_{l_{i}},$ $\bar{t}'_{l_{i}})_{i=0}^{k-1})_{l=1}^{N_{1}}, $ $VP_{U},{\mathbb{P}_{U}}\big)$.
\medskip

\noindent {\sf S} verifies: $ R\stackrel{?}{=}\frac{e(C,\tilde{g})}{e(g_{0},g)\cdot  e(g_{1},g)^{H(VP_{U})}};$
\begin{equation*}
\begin{split}
& \bar{c}\stackrel{?}{=}H\big(M_{U}^{2}||Y||\xi^{\bar{x}_{u}}g_{1}^{\bar{d}}Y^{\bar{c}}||D||g^{\bar{\alpha}}\vartheta^{\bar{\beta}}D^{\bar{c}}||\Phi||g^{\bar{\alpha'}}\vartheta^{\bar{\beta'}}\Phi^{\bar{c}}||C||R||\\
&e(\xi,g)^{\bar{x}_{u}}\cdot  e(\mathfrak{g},g)^{\bar{r}_{u}}\cdot \prod_{l=1}^{N_{1}}e(\hat{g},g)^{\bar{a}_{l}}\cdot
 \prod_{i=1}^{N_{2}}e(\eta_{i},g)^{\hat{e}'_{i}}
 \cdot e(C,g)^{-\bar{c}_{u}}\cdot  \\
 & e(\vartheta ,g)^{\bar{\alpha}'}\cdot 
  e(\vartheta,\tilde{g})^{\bar{\alpha}}\cdot  R^{\bar{c}}||Z_{1}||\cdots||Z_{N_{1}}|| g^{\bar{\gamma}_{1}}h^{\bar{a}_{1}}Z_{1}^{\bar{c}}||\cdots||\\
  & g^{\bar{\gamma}_{N_{1}}}h^{\bar{a}_{N_{1}}}Z_{N_{1}}^{\bar{c}}||B_{1_{1}}||\cdots||B_{1_{\zeta}}||
 \cdots|| B_{N_{2_{1}}}||\cdots
 || B_{N_{2_{\zeta}}}||W_{1_{1}}||\\
 &\cdots||W_{1_{\zeta}}||\cdots||W_{N_{2_{1}}}||\cdots||W_{N_{2_{\zeta}}}||e(\eta,\eta_{1})^{\hat{e}_{1}}\cdot  e(B_{1_{1}},\eta_{1})^{\hat{e}''_{1}}\cdot\\
 & W_{1_{1}}^{\bar{c}}||\cdots
  || e(\eta,\eta_{1})^{\hat{e}_{1}}\cdot e(B_{1_{\zeta}},\eta_{1})^{\hat{e}''_{1}}\cdot W_{1_{\zeta}}^{\bar{c}}||\cdots|| e(\eta,\eta_{N_{2}})^{\hat{e}_{N_{2}}}\cdot\\
  & e(B_{N_{2_{1}}},\eta_{N_{2}})^{\hat{e}''_{N_{2}}}\cdot W_{N_{2_{1}}}^{\bar{c}}||\cdots||
  e(\eta,\eta_{N_{2}})^{\hat{e}_{N_{2}}}\cdot e(B_{N_{2_{\zeta}}},\eta_{N_{2}})^{\hat{e}''_{N_{2}}}\\
  &\cdot W_{N_{2_{\zeta}}}^{\bar{c}}
 \big), \big( \bar{e}_{l}=H\big(M_{U}^{2}||Z_{l}||g_{l}^{\check{\gamma}_{l}}h^{\check{a}_{l}}(Z_{l}h^{-c_{l}})^{\bar{e}_{l}}||\\
& g^{\check{\gamma}_{l}}\prod_{i=0}^{k-1}\tilde{h}_{i}^{\bar{w}_{l_{i}}}(Z_{l}h^{-c_{l}})^{\bar{e}_{l}}||g^{\check{\gamma}_{l}}\prod_{i=0}^{k-1}\tilde{h}_{i}^{\bar{w}'_{l_{i}}}(Z_{l}h^{-d_{1}+q^{k}})^{\bar{e}_{l}}\big) \big)_{l=1}^{N_{1}},
 \end{split}
\end{equation*}
\begin{equation*}
\begin{split}
& \Big(\big(\bar{d}_{l_{i}}=H(M_{U}^{2}||A_{w_{l_{i}}}||A'_{w_{l_{i}}}||V_{{l_{i}}}||V'_{l_{i}}||e(h,h)^{\bar{t}_{l_{i}}}\cdot e(A_{w_{l_{i}}},h)^{-\hat{w}_{l_{i}}}\cdot \\
&V_{l_{i}}^{\bar{d}_{l_{i}}}|| e(h,h)^{\bar{t}'_{l_{i}}}\cdot e(A'_{w_{l_{i}}},h)^{-\hat{w}'_{l_{i}}}\cdot (V'_{l_{i}}) ^{\bar{d}_{l_{i}}}
)\big)_{i=0}^{k-1}\Big)_{l=1}^{N_{1}}\\
\end{split}
\end{equation*}

\noindent{\bf The Detail of $\prod_{U}^{3}$:}

\noindent{\sf U} selects $\pi,\lambda,\tilde{x}_{u},\tilde{s}_{u},\tilde{\pi},\tilde{\pi}',\tilde{\lambda}',\tilde{\omega}_{u},\tilde{d}_{u}\stackrel{R}{\leftarrow}\mathbb{Z}_{p}$ and $M_{U}^{3}\stackrel{R}{\leftarrow}\mathbb{G}$, and computes
\begin{equation*}
\begin{split}
& D=g^{s_{u}},\tilde{D}=g^{\tilde{s}_{u}},Ps_{U}=\xi^{x_{u}}g_{1}^{d_{u}},\tilde{Ps}_{U}=\xi^{\tilde{x}_{u}}g_{1}^{\tilde{d}_{u}},\\
&E=\xi^{x_{u}}H'(ID_{V})^{rs_{u}}, \tilde{E}=\xi^{\tilde{x}_{u}}H'(ID_{V})^{r\tilde{s}_{u}}, F=T_{U}\vartheta^{\pi},\\
&  J=g^{\pi}\vartheta^{\lambda}, \tilde{J}=g^{\tilde{\pi}}\vartheta^{\tilde{\lambda}},J'=J^{\omega_{u}}=g^{\pi\omega_{u}}\vartheta^{\lambda\omega_{u}},\\
&\tilde{J}'=J^{\tilde{\omega_{u}}}=g^{\pi\tilde{\omega}_{u}}\vartheta^{\lambda\tilde{\omega}_{u}},
R=\frac{e(F,Y_{S})}{e(g_{0},\rho)\cdot e(Ps_{U},\rho)\cdot e(g_{3},\rho)^{\psi_{u}}},\\
&~\tilde{R}=e(g_{2},\rho)^{\tilde{s}_{u}}\cdot e(F,\rho)^{-\tilde{\omega}_{u}}\cdot e(\vartheta,\rho)^{\tilde{\pi}'}\cdot e(\vartheta,Y_{S})^{\tilde{\pi}},
\end{split}
\end{equation*}
\begin{equation*}
\begin{split}
&c=H(M_{U}^{3}||D||Ps_{U}||E||J||J'||R||\tilde{D}||\tilde{Ps}_{U}||\tilde{E}||\tilde{J}||\tilde{J}'||R'),\\
& \bar{s}_{u}=\tilde{s}_{u}-cs_{u}, \bar{x}_{u}=\tilde{x}_{u}-cx_{u},\hat{s}_{u}=r\tilde{s}_{u}-crs_{u}, \bar{\pi}=\tilde{\pi}-c\pi,\\
& \bar{\lambda}=\tilde{\lambda}-c\lambda,\bar{\omega}_{u}=\tilde{\omega}_{u}-c\omega_{u}, ~\bar{\pi}'=\tilde{\pi}'-c \pi \omega_{u} ~\mbox{and}~ \bar{d}_{u}=\tilde{d}_{u}-cd_{u}.
\end{split}
\end{equation*}

\noindent {\sf U} sends $(\mathbb{P}_{U},Price,Serv,VP_{T},M_{U}^{3},D,Ps_{U},E,F,
J,J',R,c, \bar{s}_{u},$ $\bar{x}_{u},\hat{s}_{u},\bar{\pi},$ $\bar{\lambda},\bar{\omega}_{u},\bar{\pi}',\bar{d}_{u})$ to {\sf V}.
\begin{equation*}
\begin{split}
\noindent\mbox{{\sf V} verifies:} ~~
&\psi_{u}\stackrel{?}{=}H(\mathbb{P}_{U}||Price||Serv||VP_{T}),\\
&R\stackrel{?}{=}\frac{e(F,Y_{S})}{e(g_{0},\rho)\cdot e(Ps_{U},\rho)\cdot e(g_{3},\rho)^{\psi_{u}}},
\end{split}
\end{equation*}
and

\begin{equation*}
\begin{split}
& c\stackrel{?}{=}H\big(M_{U}^{3}||D||E||J||J'||R||g^{\bar{s}_{u}}D^{c}||\xi^{\bar{x}_{u}}g_{1}^{\bar{d}_{u}}Ps_{U}^{c}||\\
&\xi^{\bar{x}_{u}}H'(ID_{V})^{\hat{s}   _{u}}E^{c}||g^{\bar{\pi}}\vartheta^{\bar{\lambda}}J^{c} ||J^{\bar{\omega}_{u}}J'^{c}||e(g_{2},\rho)^{\bar{s}_{u}}\cdot  \\
&  e(F,\rho)^{-\bar{\omega}_{u}}\cdot e(\vartheta,\rho)^{\bar{\pi}'}\cdot e(\vartheta,Y_{S})^{\bar{\pi}}R^{c}\big).
\end{split}
\end{equation*}
\end{document}